\documentclass[12pt,pdftex]{article}  %%

\usepackage[top=3.5cm,bottom=3.5cm,left=2cm,right=2cm]{geometry}
\usepackage{amssymb,amsmath}
\usepackage{slashed}
\usepackage{hyperref}
\usepackage{cite}
\usepackage{graphicx}%%%
\usepackage[title]{appendix}
\numberwithin{equation}{section}
\begin{document}
\title{
\vbox{
\baselineskip 14pt
\hfill \hbox{\normalsize KUNS-2677
}} \vskip 1cm
\bf \Large Axial U(1) current in Grabowska and Kaplan's formulation
\vskip 0.5cm
}
\author{
Yu~Hamada\thanks{E-mail: \tt yu.hamada@gauge.scphys.kyoto-u.ac.jp}~ and
Hikaru~Kawai\thanks{E-mail: \tt hkawai@gauge.scphys.kyoto-u.ac.jp}
\bigskip\\
%\\*[20pt]
\it \normalsize
 Department of Physics, Kyoto University, Kyoto 606-8502, Japan\\
\smallskip
}
\date{}

\setcounter{page}{0}
\maketitle\thispagestyle{empty}
\abstract{\normalsize
 Recently, Grabowska and Kaplan suggested a non-perturbative formulation of a chiral gauge theory, which consists of the conventional domain-wall fermion and a gauge field that evolves by the gradient flow from one domain wall to the other.
 %A long distance flow makes the gauge field pure gauge (or instanton) and one of the chiral modes of the domain-wall fermion (fluffy mirror fermion) to decouple.
 %Therefore, it can be expected to obtain non-perturbative a chiral gauge theory.
 
 In this paper, we discuss the U(1) axial-vector current in 4 dimensions using this formulation.
 %anomaly in a 4-dimensional vector-like gauge theory using this 5-dimensional formulation.
 We introduce two sets of domain-wall fermions belonging to complex conjugate representations so that the effective theory is a 4-dimensional vector-like gauge theory.
 Then, as a natural definition of the axial-vector current, we consider a current that generates the simultaneous phase transformations for the massless modes in 4 dimensions.
 However, this current is exactly conserved and does not reproduce the correct anomaly.
 In order to investigate this point precisely, we consider the mechanism of the conservation. 
% we investigate the conserving mechanism in terms of calculating an effective action.
 We find that this current includes not only the axial current on the domain wall but also a contribution from the bulk, which is non-local in the sense of 4-dimensional fields.
 Therefore, the local current is obtained by subtracting the bulk contribution from it.
}
\newpage
\section{Introduction}\label{sec:introduction}
\quad
 Formulating a chiral gauge theory non-perturbatively has been a long-standing problem\cite{Kaplan:1992bt,Neuberger:1997fp,Narayanan:1994gw,Jansen:1994ym,Shamir:1993zy,Furman:1994ky,Aoki:1993rg,Narayanan:1992wx,Golterman:1992ub,Luscher:1999un}.
% In particular, there have been many attempts using the domain-wall fermion\cite{,
 Recently\cite{Grabowska:2015qpk,Grabowska:2016bis}, Grabowska and Kaplan suggested a formulation that consists of the domain-wall fermion in $2n$+1 dimensions and a gauge field that evolves by the gradient flow from one domain wall to the other.
 A long-distance flow makes the gauge field pure gauge, and thus one of the massless modes (``fluffy mirror fermion'' or ``fluff'') does not couple with the gauge field.
 Therefore, we obtain a chiral gauge theory including only the other massless mode that couples with the gauge field.
 However, the heavy modes in the bulk induce some terms, which can not be renormalized to the 4-dimensional Lagrangian.
 %include not only the massless mode and the gauge field but also non-local terms induced by the heavy modes of the domain-wall fermion in the bulk.
% These non-local terms are obstacles to regard this effective theory as a $2n$-dimensional local theory.
 To cancel the bulk terms, Grabowska and Kaplan introduced a subtracting field, which has a loop factor +1 and a constant mass.
 %This field plays a role 
 It is known that the cancellation is not complete, but there remains a Chern-Simons-like term\cite{Callan:1984sa,Deser:1981wh}.
 However, if the anomaly-free condition $d^{abc}=0$ is satisfied, the Chern-Simons-like term vanishes and then we obtain the 4-dimensional local theory. 
 
 % The formulation in Ref~\cite{Grabowska:2015qpk,Grabowska:2016bis} looks well.
%  One way to study this formulation is introducing two sets of domain-wall fermions belonging to complex conjugate representations in 5 dimensions.
%to simulate a $2n$-dimensional vector-like gauge theory by
 In order to investigate the consistency of this formulation, we consider a vector-like theory by introducing two sets of domain-wall fermions belonging to complex conjugate representations\cite{Grabowska:2016bis,Suzuki:2016,Okumura:2016dsr,Makino:2016auf}.
 %to simulate a $2n$-dimensional vector-like gauge theory by
 Each of the fermions induces one left-handed physical fermion and one right-handed fluff fermion.
% The effective theory for this system consists of two left-handed physical fermions belonging to each representation and two right-handed fluff fermions on each domain wall.
 Therefore, if the fluffs are decoupled correctly, we have the 4-dimensional vector-like gauge theory with one right-handed and one left-handed chiral fermions after we apply the charge conjugation to one of the physical fermions.   
 % However, it is non-trivial how to define a current in the $2n$+1-dimensional theory which behaves as an anomalous current in the $2n$-dimensional theory.
% However, if one defines the axial-vector current naively in 5 dimensions, it does not reproduce the correct anomaly but is conserved\cite{Suzuki:2016}.
% In this paper, we consider a 5-dimensional U(1) gauge theory and two sets of domain-wall fermions with charge $\pm 1$.
 In this paper, we consider the U(1) axial-vector current and discuss how the anomaly arises.
 %anomaly in a 4-dimensional vector-like gauge theory using this 5-dimensional formulation.
% We introduce two sets of domain-wall fermions belonging to complex conjugate representations so that the effective theory is a 4-dimensional vector-like gauge theory.
  %As a naive candidate of the axial-vector current,
  We first define a current that generates the simultaneous phase transformations for the left-handed physical fermions in 4 dimensions.
  From the viewpoint of the effective theory, this current looks like the U(1) axial-vector current.
%In the effective theory, this current generates the chiral transformation of the chiral fermions
  However, it is pointed out in Ref.\cite{Suzuki:2016} that this current is exactly conserved and does not reproduce the correct anomaly.
  In order to solve this paradox, we investigate the mechanism of the conservation.
  We find that this current contains a bulk contribution in addition to the axial-vector current on the domain wall.
  Therefore, the proper local current is obtained by subtracting the bulk part.
% However, this current is exactly conserved and does not reproduce the correct anomaly.
% So we investigate the conserving mechanism in terms of calculating an effective action.
% We find that this 
% Indeed this current is found to be conserved exactly, but this is because this current includes not only an axial current but also a Chern-Simons current, although the anomaly-free condition is satisfied.
% This current is nothing but the gauge variant but conserved current in QCD U(1) problem\cite{tHooft:1986ooh}.
%
% However, it is not well understood beyond the naive argument.
% We begin with n more deeply in the continuum.
% In Ref~\cite{Grabowska:2015qpk,Grabowska:2016bis}, however, lattice regularization is assumed.

 This paper is organized as follows.
 In Sec.~\ref{sec:revi-grab-kapl}, we review the formulation of Grabowska and Kaplan in the lattice space.
 In Sec.~\ref{sec:regul-form}, we consider a regularization of this formulation in the continuum space in order to simplify the calculations in the subsequent sections.
 We find that one needs to introduce Pauli-Villars fields for both of the domain-wall fermion and the subtracting field.
% This regularization seems to generate some problems, but these will be eliminated.
  In Sec.~\ref{sec:calc-doma-wall}, we calculate the 1-loop effective action to the quadratic order in the gauge fields, and confirm that the effective action consists of three parts.
  One is equal to the effective action of a chiral fermion with Pauli-Villars-like regularization.
  The second is the Chern-Simons term in the bulk.
  The third are various divergent terms, which will be cancelled by our regularization.
% In addition, we make sure that the problems appearing in Sec.~\ref{sec:regul-form} can be eliminated.
 In Sec.~\ref{sec:constr-u1-axial}, we discuss the axial-vector current in 4-dimensions.
 In Sec.~\ref{sec:summary-conclusions}, we give summary and conclusions.
% Consequently, we have to subtract the Chern-Simons current from the naively defined current.
% This current looks like a U(1) axial current, but can not be anomalous because of gauge invariance.
% 
% 
%
\section{Review of Grabowska and Kaplan's method}\label{sec:revi-grab-kapl}
\quad
We review the formulation of Grabowska and Kaplan.
There are recent studies\cite{Okumura:2016dsr,Makino:2016auf,Fukaya:2016dcl} based on this formulation.
In this section, we consider the lattice space although we use the symbols in the continuum space.
We will discuss the continuum regularization in the next section.

We start with a domain-wall fermion in $2n$+1 dimensions:
\begin{equation}
  \mathcal{L} = \bar{\psi} [\slashed{\partial}_{2n+1} - M \epsilon(s) ]\psi .
\end{equation}
Here $\psi $ is the Dirac field with $2^n$ components.
 $s$ is the $2n$+1th coordinate, $s\in [-L,L]$ with periodic boundary condition, and $\epsilon (s)=\mathrm{sgn}(s)$.
In the limit of $L\rightarrow\infty$, two massless modes are localized on the $2n$-dimensional wall $s=0$ and $s=L$, which have the chirality $-1$ and $+1$ respectively.
The heavy modes that live in the bulk will be decoupled classically in the limit of $M\rightarrow \infty$.
In order to obtain a chiral gauge theory, in which only the left-handed mode couples with the gauge field, the $2n$+1-dimensional gauge field $\bar{A}_\mu$ is constructed by the gradient flow\cite{Narayanan:2006rf,Luscher:2010iy,Luscher:2011bx} from $s=0$ to $s=\pm L$:
\begin{equation}
 \partial_s \bar{A}_\nu(x,s)=\frac{\epsilon(s)}{M'}D_\mu\bar{F}_{\mu\nu}~,\label{eq:91} 
\end{equation}
with $\bar{A}_\mu(x,0)=A_\mu(x)$, $\mu,\nu=1,\cdots,2n$, and $\bar{A}_{2n+1}=0$.
$\bar{F}_{\mu\nu}$ is the field strength of the gauge field $\bar{A}_\mu$.
%One of the features of the gradient-flow is that the gauge field becomes non-local.
We assume $M' \gg M$ so that $\bar{A}_\mu(x,s)$ is close to $A_\mu(x)$ near the domain wall $|s| \lesssim 1/M$.
Since the gradient flow damps the physical degrees of freedom, the gauge field $\bar{A}_\mu$ becomes to pure gauge\footnote{More precisely, it is also possible for the gauge field to become an instanton configuration. We don't consider this case in this paper.} at $s=L$ in the limit of $L\rightarrow\infty$.
Thus the right-handed mode on $s=L$ is decoupled and we obtain the $2n$-dimensional chiral gauge theory if the bulk degrees of freedom are decoupled.

In order to cancel the bulk degrees of freedom\footnote{For the case that the gauge field is constant in the $s$ direction, $\bar{A}_\mu(x,s) = A_\mu(x)$, we don't have to cancel the bulk terms because they can be absorbed in the 4-dimensional Lagrangian. However, it is not possible when we consider the gradient flow.}, we introduce a ``subtracting field''\footnote{In Ref.\cite{Grabowska:2015qpk,Grabowska:2016bis}, this field is called "Pauli-Villars field". But we distinguish this field from a conventional Pauli-Villars field, whose role is a regularization.}, which has a loop factor +1 and a constant mass $-M$.
%In the bulk far from the domain walls, the domain-wall fermion seems to have a constant mass $\pm M$ and massless modes are negligible.
This setting is equivalent to defining the fermion determinant as follows:
\begin{equation}
 \Delta (A)\equiv\frac{\mathrm{det}\left( \slashed{D}^{(R)}_{2n+1} - M \epsilon(s)\right) }{\mathrm{det}\left( \slashed{D}^{(R)}_{2n+1} +M \right) } , \label{fermion_det}
\end{equation}
where $\slashed{D}^{(R)}_{2n+1}$ is the $2n$+1-dimensional Dirac operator belonging to the representation $R$.
%
%Since the heavy mode in the bulk looks like a fermion with a constant mass $+M$ or $-M$, the non-local operators induced by the heavy modes is cancelled with that induced by the subtracting field, at least naively.
Indeed the terms that are even functions of $M$ in the bulk are cancelled.
On the other hand, for the odd terms, the parity anomaly survives and the effective action contains a bulk term\cite{Callan:1984sa,Deser:1981wh}:
\begin{equation}
 S^{(CS)}_{2n+1} = c_{2n+1} \frac{M}{|M|}\int [\epsilon(s) + 1 ]~\omega_{2n+1} \label{eq:35}.
\end{equation}
Here
\begin{equation}
  c_{2n+1}=\frac{i^n}{2^{n+1}\pi^n(n+1)!},
\end{equation}
and $\omega_{2n+1}$ is the $2n$+1-dimensional Chern-Simons form.
$S_{2n+1}^{(CS)}$ vanishes if the representation satisfies the condition for the anomaly cancellation in $2n$ dimensions.

%This formulation looks successful naively.
%It is convenient to consider the continuum theory for 
In order to perform the calculation easily, we consider a continuum version of this formulation in the following sections.
%
% 
%
%In our convention, $\gamma^5$ means always the chirality matrix, i,e, $\gamma^5\equiv\gamma^1 \cdots \gamma^{2n}$ even for $2n$+1-dimensional theory, unless it is confusing.
%%And $a=1/|M|$ is lattice spacing.
%The numerator in Eq.(\ref{fermion_det}) is a domain-wall fermion determinant and the denominator is Pauli-Villars-like one.
%Note that this Pauli-Villars-like field is used to subtract bulk heavy modes of the domain-wall fermion, not to regulate the formulation.
%%Regularization is automatically achieved in lattice theory.
%%It is thought naively that heavy modes of the domain-wall fermion with mass $M$ in bulk is cancelled by Pauli-Villars field and
%Here, a Chern-Simons term induced in the bulk is non-local due to the gradient flow and does harm to the effective theory.
%%Therefore, one can obtain an effective theory consisting of a chiral fermion coupling to physical gauge field at $s=0$ and non-local a Chern-Simons term propotional $d^{abc}$.
%
%
%
%
\section{Regularization in the continuum formulation}\label{sec:regul-form}
 \quad
In this section, we regularize the formulation given in Sec.~\ref{sec:revi-grab-kapl} in the continuum space.
The bare effective action corresponding to Eq.~(\ref{fermion_det}) is given by\footnote{We drop the superscript "$(R)$" in $\slashed{D}_{2n+1}$.}
\begin{equation}
  \log \Delta (A) = \mathrm{Tr}\log\left( \slashed{D}_{2n+1} - M \epsilon(s)\right)- \mathrm{Tr}\log\left( \slashed{D}_{2n+1} + M \right) \label{eq:2}.
\end{equation}
 %, we begin with regularizing this formulation by Pauli-Villars regularization for $n=2$.
%Since Eq.~(\ref{eq:2}) includes the chirality matrix $\gamma^5$,
Here, we adopt the Pauli-Villars regularization\footnote{The dimensional regularization can not be used for the $2^n$-component Dirac field in $2n+1$ dimensions.}.
%We consider only a case $n=2$ for simplicity, but the following argument can be extended easily to general cases.
%The below discussion can be applied to general $n$.
%
%In the present paper, we call the field with constant mass in Eq.~(\ref{eq:2}) ``subtracting field'', introduced to subtract the heavy modes contribution in the bulk, and we distinguish this field from Pauli-Villars fields which we will introduce to regulate. 

We regularize the domain-wall fermion and the subtracting field respectively as follows:

\begin{equation}
  \mathrm{Tr}\log\left( \slashed{D}_{2n+1} - M \epsilon(s)\right) \rightarrow \mathrm{Tr}\log\left( \slashed{D}_{2n+1} - M \epsilon(s)\right) + \sum_i C_i \mathrm{Tr}\log\left( \slashed{D}_{2n+1} - M_i \epsilon(s)\right),\label{eq:53}
\end{equation}
\begin{equation}
  \mathrm{Tr}\log\left( \slashed{D}_{2n+1} + M \right) \rightarrow \mathrm{Tr}\log\left( \slashed{D}_{2n+1} + M \right) + \sum_i C_i' \mathrm{Tr}\log\left( \slashed{D}_{2n+1} + M_i' \right).\label{eq:54}
\end{equation}
Note that while the subtracting field is regularized as usual, the domain-wall fermion is regularized by additional domain-wall fermions with mass $M_i\epsilon(s)$.
%, which we call ``domain-wall Pauli-Villars fields'' .
The parameters $C_i,M_i,C_i',M_i'$ will be determined later so that the regularized effective action converges as usual.
Here, we choose $C_i'=C_i$ and $M_i'=M_i$ so that the Pauli-Villars fields do not generate extra bulk effective action.
In other words, we introduce pairs of Pauli-Villars fields consisting of a domain-wall fermion and a subtracting field, which we call Pauli-Villars pairs.
Thus the regularized effective action is
\begin{eqnarray}
 \log \Delta (A)_\mathrm{reg.} &=& \mathrm{Tr}\log\left( \slashed{D}_{2n+1} - M \epsilon(s)\right)- \mathrm{Tr}\log\left( \slashed{D}_{2n+1} + M \right) \nonumber \\
 && +\sum_i C_i \left[ \mathrm{Tr}\log\left( \slashed{D}_{2n+1} - M_i \epsilon(s)\right) - \mathrm{Tr}\log\left( \slashed{D}_{2n+1} + M_i \right) \right] . \label{fermion_det_reg5}
\end{eqnarray}
%We investigate the condition of the parameters. 
%In a general 5-dimensional theory, the condition to regularize a fermion loop with mass $M_0$ by Pauli-Villars is

Let us write down the condition for the effective action to converge.
For a necessary condition, divergences arising on the walls should be cancelled.
As we will see in Eq.~(\ref{eq:44}), a pair of a domain-wall fermion and a subtracting field behaves like a chiral fermion with a Pauli-Villars-like field\footnote{This $2n$-dimensional Pauli-Villars-like field is not the Pauli-Villars field that we have introduced in Eq.~(\ref{eq:53}) and Eq.~(\ref{eq:54}).} around $s=0$.
Therefore, all pairs including the Pauli-Villars pairs give the following contribution to the effective action from the near-wall region:
\begin{eqnarray}
&& \mathrm{Tr}\log\left( \slashed{D}_{2n+1} - M \epsilon(s)\right)- \mathrm{Tr}\log\left( \slashed{D}_{2n+1} + M \right) \nonumber \\
 &&\hspace{1em} +\sum_{i} C_i \left[ \mathrm{Tr}\log\left( \slashed{D}_{2n+1} - M_i \epsilon(s)\right) - \mathrm{Tr}\log\left( \slashed{D}_{2n+1} + M_i \right) \right] \nonumber \\
 &&\xrightarrow{\mathrm{around}~s=0} \left[ \mathrm{Tr}\log (\slashed{D}_{2n}P_- + \slashed{\partial}_{2n}P_+ ) - \mathrm{Tr}\log (\slashed{D}_{2n}P_- + \slashed{\partial}_{2n}P_+  -M) \right] \nonumber \\
 &&\hspace{4em} +\sum_{i} C_i \left[ \mathrm{Tr}\log (\slashed{D}_{2n}P_- + \slashed{\partial}_{2n}P_+ ) - \mathrm{Tr}\log (\slashed{D}_{2n}P_- + \slashed{\partial}_{2n}P_+  -M_i) \right] , \label{eq:48}
\end{eqnarray}
where $P_-$ and $P_+$ are the chirality projection operators.
 $\mathrm{Tr}\log(\slashed{D}_{2n}P_- + \slashed{\partial}_{2n}P_+ )$ and  $ \mathrm{Tr}\log (\slashed{D}_{2n}P_- + \slashed{\partial}_{2n}P_+  -M)$ are the effective action of the left-handed chiral fermion and the Pauli-Villars-like field, respectively.
We will derive Eq.~(\ref{eq:48}) in Sec.~\ref{sec:calc-doma-wall}.
The conditions to cancel the divergences in Eq.~(\ref{eq:48}) are\footnote{Generally, we need $d$ conditions in $d$ dimensions. }
\begin{eqnarray}
 \begin{aligned}
 M + \sum_i C_i M_i &=&0 , \\
 M^2 + \sum_i C_i (M_i)^2 &=&0 , \\
 M^3 + \sum_i C_i (M_i)^3 &=&0 , \\
  \vdots&&.
   \end{aligned}\label{eq:50}
\end{eqnarray}
Note that the leading divergences in Eq.~(\ref{eq:48}), which are independent of $M$ and $M_i$, are cancelled in each pair.

Eq.~(\ref{eq:50}) are also sufficient to cancel the divergences from the bulk.
%because divergences arising on the walls still need to be cancelled.
In the bulk region $-L<s<0$, the cancellation is trivial because the domain-wall fermions and the subtracting fields have the same mass in each pair.
%We assume that it is sufficient to regularize the loops of the heavy modes and the subtracting fields in the bulk.
%So we consider only the bulk $-L<s<0$ and $0<s<L$.
In the bulk region $0<s<L$, Eq.~(\ref{fermion_det_reg5}) reduces to
\begin{eqnarray}
&& \mathrm{Tr}\log\left( \slashed{D}_{2n+1} - M \right)- \mathrm{Tr}\log\left( \slashed{D}_{2n+1} + M \right) \nonumber \\
&& \hspace{2em} +\sum_i C_i \left[ \mathrm{Tr}\log\left( \slashed{D}_{2n+1} - M_i \right) - \mathrm{Tr}\log\left( \slashed{D}_{2n+1} + M_i \right) \right] . \label{eq:3}
\end{eqnarray}
% $M_i\epsilon(s) \rightarrow M_i$.
In Eq.~(\ref{eq:3}), terms that are even functions of $M$ and $M_i$ are trivially cancelled.
On the other hand, the odd terms are cancelled if the following conditions are satisfied:
\begin{eqnarray}
 \begin{aligned}
 M + \sum_i C_i M_i &=& 0 , \\
 M^3+\sum_i C_i (M_i)^3 &=& 0 , \\
  \vdots& &    ,
 \end{aligned} \label{eq:52} 
\end{eqnarray}
which are part of Eq.~(\ref{eq:50}).
%because divergences arising on the walls are automatically cancelled.
%Indeed, as we will see in Eq.~(\ref{eq:22}), the subtracting fields generate Pauli-Villars fields on the wall that regularize the massless modes.
%The second problem is that the domain-wall Pauli-Villars fields become massless on the domain walls $s=0,L$.
%So one can not use these fields to regularize the loops of the original domain-wall fermion on the walls.
%However, this problem is solved 
%Note that Eq.~(\ref{eq:4}) is sufficient 
%Therefore, we adopt Eq.~(\ref{eq:4}) as the condition to regularize the effective action.
%It is sufficient to consider only the one side of the bulk $-L<s<0$ because $M\epsilon(s) = M$ in the bulk $0<s<L$ and each pairs completely cancel there.
%And the contribution from the bulk $0<s<L$ vanishes.% since $M\epsilon(s)\rightarrow M$.
Therefore, Eq.~(\ref{eq:50}) is the necessary and sufficient condition for the effective action to converge.

However, we need to prevent the Pauli-Villars fields from changing the physical degrees of freedom.
In fact, each of the Pauli-Villars pairs induces a massless mode on the wall and a Chern-Simons term in the bulk, which will not be decoupled even if we take the limit $M_i\rightarrow\infty$.
Thus one observes $\sum_i C_i$ additional massless modes and Chern-Simons terms.
%However, we can avoid this problem as follows.
%As we mentioned from Eq.~(\ref{eq:48}), the massless modes contribute to the effective action as chiral fermions, i,e, $\mathrm{Tr}\log(\slashed{D}_{2n}P_- + \slashed{\partial}_{2n}P_+ )$.
%\footnote{$P_-$ and $P_+$ are the chirality projection operators.}
%$S_{2n+1}^{(CS)}$.
%In addition, the Chern-Simons term $S_{2n+1}^{(CS)}$ is independent of the mass.
%Therefore, the effective action of the chiral fermions and the Chern-Simons terms, togther with the extra contributions, are respectively
%\begin{equation}
% \left(1+ \sum_i C_i \right)  \mathrm{Tr}\log(\slashed{D}_{2n}P_- + \slashed{\partial}_{2n}P_+ ) ,
%\end{equation}
%\begin{equation}
% \left( 1+ \sum_i C_i \right) S_{2n+1} ^{(CS)}.
%\end{equation}
These extra contributions vanish by imposing an additional condition:
\begin{equation}
  \sum_i C_i=0 , \label{eq:45}
\end{equation}
%As we will make sure in Sec.~\ref{sec:comp-with-chir}, this effective action consists of two parts.
%One is the sum of effective actions of chiral fermions, such as $\mathrm{Tr}\log(\slashed{D}_{2n}P_- + \slashed{\partial}_{2n}P_+ )$ and $\mathrm{Tr}\log(\slashed{D}_{2n}P_+)$.\footnote{$P_-$ and $P_+$ are the chirality projection operators.}
%
%This makes such extra chiral fermions and Chern-Simons terms vanish,
%Then they will not be observed.
which we will confirm in Eq.~(\ref{eq:22}).

Thus we conclude that a continuum version of the regularized effective action is given by Eq.~(\ref{fermion_det_reg5}) with Eq.~(\ref{eq:50}) and Eq.~(\ref{eq:45}).

%This argument can be extended easily to general $2n$+1-dimensions\footnote{Note that for $n=1$, Pauli-Villars pairs are needless.}.
%
 \section{Calculation of the effective action}\label{sec:calc-doma-wall}
 \quad
In this section, we calculate the regularized effective action, Eq.~(\ref{fermion_det_reg5}), by expanding with respect to the gauge field $\bar{A}_\mu$.
 In order to do it, it is sufficient to calculate the one pair of a domain-wall fermion and a subtracting field:
 \begin{equation}
 \mathrm{Tr}\log (\slashed{D}_{2n+1}- M \epsilon(s) ) - \mathrm{Tr}\log (\slashed{D}_{2n+1} + M). \label{eq:18}
 \end{equation}
 The other pairs are obtained by replacing the mass and loop factor.
  As we will see later, Eq.~(\ref{eq:18}) consists of three parts.
 One is the effective action of the $2n$-dimensional chiral fermions with a Pauli-Villars-like regularization.
 This confirms that the massless modes localized on the walls behave as chiral fermions even at the quantum level.
 The second is the Chern-Simons term in $2n$+1 dimensions.
 The third are various divergent terms, which will be cancelled after summing up with the Pauli-Villars pairs.
 \subsection{Propagator of domain-wall fermion}\label{sec:deriv-prop-doma}
 \quad
 We begin with deriving the propagator of the domain-wall fermion in the continuum\footnote{The propagator in the lattice theory is derived in Ref\cite{Aoki:1993rg,Narayanan:1992wx} .}.
As we will see below, this propagator can be regarded as a sum of two processes.
 One is the bulk propagation with a constant mass $\pm M$.
 The other is the massless propagation along the domain walls.
 Thus this propagator includes both of the heavy bulk modes and the massless domain-wall modes.
 
The propagator is a solution of the following equation:
\begin{equation}
 \left[i \slashed{p}+\gamma^5\partial_s - \epsilon(s)M \right]G(p,s;s')=\delta(s-s'), \label{green_eq}
\end{equation}
where $G(p,s;s')$ is the Fourier transform of the propagator in the $2n$ directions,
 \begin{equation}
  G(x,s;x',s') = \int \frac{d^{2n}p}{(2\pi)^{2n}} ~e^{-ip\cdot (x-x')} ~G(p,s;s') .
 \end{equation}
We use the symbol $\gamma^5$ as the chirality matrix even in $2n$+1 dimensions, i,e, $\gamma^5\equiv\gamma^1\cdots\gamma^{2n}$.
%We consider for $s'>0$.

In order to concentrate on the modes around $s=0$, we take the limit $L\rightarrow\infty$, and obtain the following expression (See Appendix \ref{appendix}):
 \begin{equation} 
G(p,s;s')= \begin{cases}
	    S^{(+)} (p,s-s') ~+ ~\mathcal{D}^{(+)}(p) ~ e^{-(s'+s)\sqrt{p^2+M^2}}  &(0<s,s')\\
	    S^{(-)} (p,s-s') ~+ ~\mathcal{D}^{(-)}(p) ~ e^{(s'+s)\sqrt{p^2+M^2}}  &(s,s'<0)\\
	   \mathcal{D}^{(-+)}(p) ~ e^{(s-s')\sqrt{p^2+M^2}} &(s<0<s')\\
	    \mathcal{D}^{(+-)}(p) ~ e^{(s'-s)\sqrt{p^2+M^2}} &(s'<0<s) ,
	    \end{cases}\label{eq:29}
 \end{equation}
where
\begin{eqnarray} 
 S^{(+)} (p,s-s') &=& - \theta(s-s') ~ \frac{i\slashed{p}+M-\sqrt{p^2+M^2}\gamma^5}{2\sqrt{p^2+M^2}} ~ e^{(s'-s)\sqrt{p^2+M^2}} \nonumber \\
 &&  -\theta(s'-s) ~ \frac{i\slashed{p}+M+\sqrt{p^2+M^2}\gamma^5}{2\sqrt{p^2+M^2}} ~ e^{(s-s')\sqrt{p^2+M^2}} , \label{eq:10}
% S^{(-)} (p,s-s') &=& - \theta(s-s') ~ \frac{i\slashed{p}-M-\sqrt{p^2+M^2}\gamma^5}{2\sqrt{p^2+M^2}} ~ e^{(s'-s)\sqrt{p^2+M^2}} \nonumber \\
 % &&  -\theta(s'-s) ~ \frac{i\slashed{p}-M+\sqrt{p^2+M^2}\gamma^5}{2\sqrt{p^2+M^2}} ~ e^{(s-s')\sqrt{p^2+M^2}}  \label{eq:39}
 \end{eqnarray}
\begin{equation}
  \mathcal{D}^{(+)}(p)= - \frac{i\slashed{p}M(M + \sqrt{p^2+M^2} \gamma^5)}{2p^2\sqrt{p^2+M^2}} +\frac{M}{2\sqrt{p^2+M^2}}, \label{eq:40} 
\end{equation} 
\begin{equation}
 \mathcal{D}^{(-+)}(p)= - \frac{i\slashed{p} (\sqrt{p^2+M^2}+M\gamma^5)}{2p^2} -\frac{1}{2}\gamma^5 \label{eq:41} ,
\end{equation}
and $\theta(s-s')$ is the step function.
 $S^{(-)}$ is obtained from replacing $M\rightarrow -M$ in $S^{(+)}$.
 $\mathcal{D}^{(-)}, \mathcal{D}^{(+-)}$ are obtained from $\mathcal{D}^{(+)}, \mathcal{D}^{(-+)}$, respectively, by replacing $M\rightarrow -M $ and $ \gamma^5\rightarrow-\gamma^5$.
%$\theta(s-s')$ is a step function.
%Concrete expressions of $\mathcal{D}^{(+)},\mathcal{D}^{(-)},\mathcal{D}^{(-+)},\mathcal{D}^{(+-)}$ are given in Appendix~\ref{appendix}. 
Note that $S^{(+)}$ and $S^{(-)}$ are the conventional propagators in $2n$+1 dimensions with constant mass $M$ and $-M$, respectively, and represent the heavy modes.
The other terms in Eq.~(\ref{eq:29}) represent the massless modes localized on the wall $s=0$.

These results are consistent with the physical intuition that the propagator $G(p,s;s')$ reduces to the conventional one with constant mass $\pm M$ in the region far from the domain wall, $s,s'\ll -1/M$ or $1/M \ll s,s'$.
\subsection{Calculation of effective action   --vacuum polarization-- }\label{sec:culc-effect-acti}
\quad
Let us expand Eq.~(\ref{eq:18}) as follows:
\begin{eqnarray}
 && \mathrm{Tr}\log (\slashed{D}_{2n+1}-\epsilon(s)M) - \mathrm{Tr}\log (\slashed{D}_{2n+1}+M) \label{eq:68} \\
&=& \sum_m \frac{1}{m}~\left( \prod_i^m \int \frac{d^dk_i}{(2\pi)^d} \right) \left( \prod_i^m \int_{-L}^L ds_i \right) (2\pi)^d~\delta^{(d)}(\sum_i k_i) \nonumber \\
 &&\hspace{2em} \mathrm{tr} \left[\bar{A}_{\mu_1}(k_1,s_1)\cdots \bar{A}_{\mu_m}(k_m,s_m)\right]~\Gamma^{\mu_1\cdots\mu_m}(\{k_i\},\{s_i\}) \label{eq:9} \\
 &\equiv& \sum _m \frac{1}{m} ~ I_m^{(L)} ~, \label{eq:67}
\end{eqnarray}
where $d=2n$ and $\Gamma^{\mu_1 \cdots\mu_m}$ is the sum of the fermion loops with $m$ vertices for the domain-wall fermion and the subtracting field.
Note that $k_i(i = 1,\cdots,m)$ are the $2n$-dimensional momenta, and $s_i$ are the $2n$+1th coordinates.
As in the previous section, we take the limit $L\rightarrow\infty$ and consider
\begin{equation}
 \lim _{L \to \infty} I_m^{(L)} \equiv I_m \label{eq:56}.
\end{equation}

%As we have mentioned ve, our statement is that Eq.~(\ref{eq:9}) consists of two parts.
%One is equal to the effective action of a chiral fermion with its Pauli-Villars field on the wall .
%The other is the Chern-Simons term in $2n$+1 dimensions.
%We do not confirm this statement in all orders of the perturbation theory.
%Instead, we calculate only the leading term in the above expansion:
In the following, we give an explicit calculation for $I_2$, which is nothing but the vacuum polarization loop:
\begin{equation}
I_2 = \int \frac{d^dk}{(2\pi)^d}  ~ \int _{-\infty}^\infty ds~\int_{-\infty}^\infty ds' ~ \mathrm{tr} \left[\bar{A}_\mu(-k,s') \bar{A}_\nu(k,s) \right] \Gamma^{\mu\nu}(k,s,s'),\label{eq:7}
\end{equation}
where
\begin{equation}
 \Gamma^{\mu\nu}(k,s,s')= \int \frac{d^dp}{(2\pi)^d} \left[ \mathrm{tr}\left[ \gamma^\mu G(p,s';s) \gamma^\nu G(p',s;s') \right] -  \mathrm{tr}\left[ \gamma^\mu S^{(-)}(p,s';s) \gamma^\nu S^{(-)}(p',s;s') \right] \right]. \label{eq:42}
\end{equation}
 Here $p'$ stands for $p+k$ so that we must substitute $p'=p+k$ in Eq.~(\ref{eq:42}) before integrating with respect to $p$.
 The second term in Eq.~(\ref{eq:42}) comes from the subtracting field.
% Note that we have taken the limit $L\rightarrow\infty$ as the previous section.

\begin{figure}[htbp]
 \begin{center}
  \includegraphics[width=0.5\textwidth]{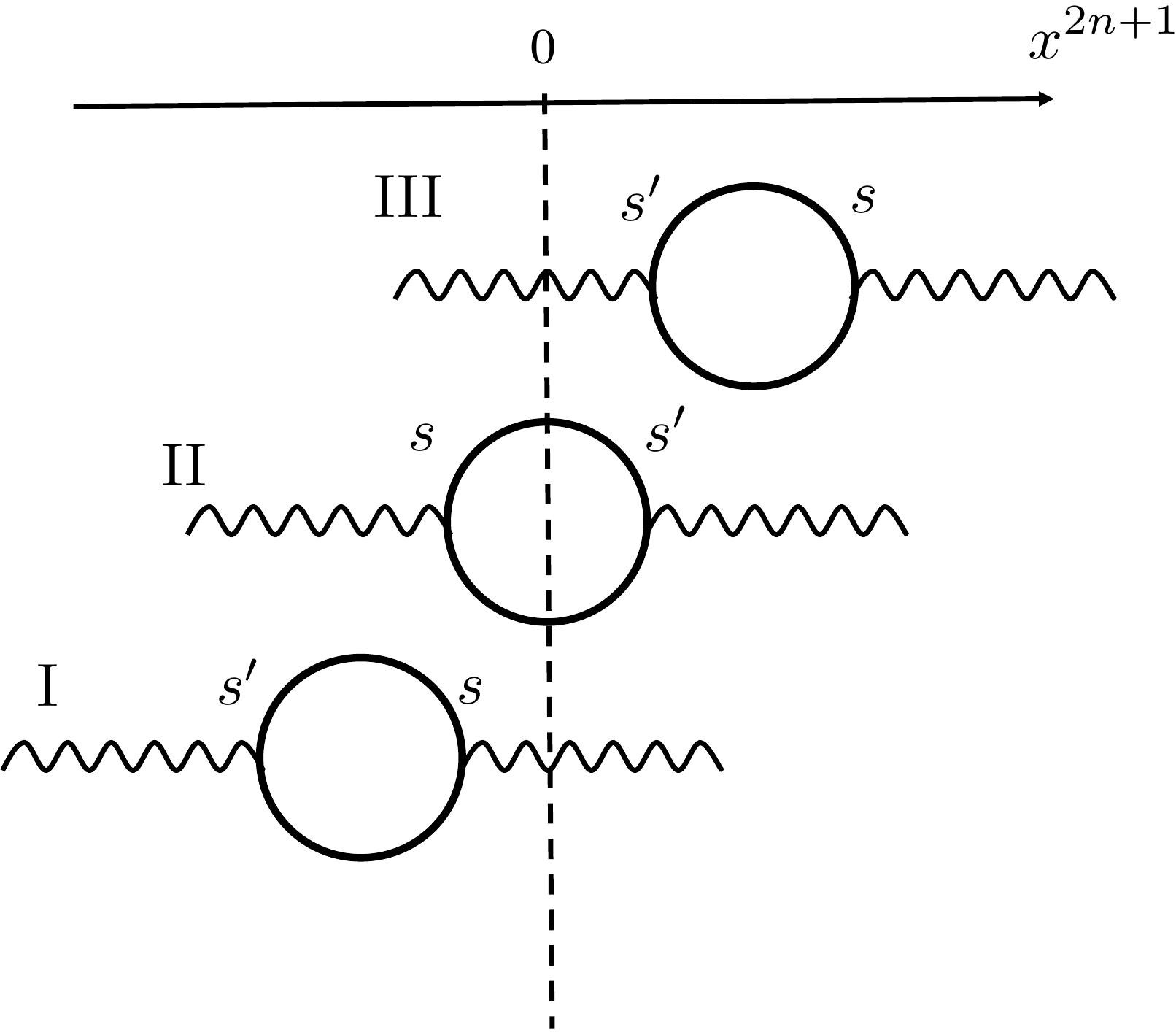}
  \caption{Feynman diagrams for three regions I,II,III\label{fig:1}}
 \end{center}
\end{figure}

It is convenient to divide the range of $s$ and $s'$ into six regions:
\begin{equation}
\mathrm{I}:\{s'<s<0\} \oplus \mathrm{II}:\{s<0<s'\} \oplus \mathrm{III}:\{0<s'<s\} \oplus \mathrm{I'} \oplus \mathrm{II'} \oplus \mathrm{III'} , \label{eq:46}
\end{equation}
where the regions I, II, III correspond to diagrams in Fig.~\ref{fig:1}.
 $\mathrm{I'},~ \mathrm{II'},~ \mathrm{III'}$ are obtained by interchanging $s\leftrightarrow s'$ in $\mathrm{I},~\mathrm{II},~\mathrm{III}$, respectively.
%We call the first three terms in Eq.~(\ref{eq:46}) ``region I, II, III''. 

We denote the contribution from the region I by I:
\begin{equation}
\mathrm{I}\equiv \iint _{(\mathrm{I})} ~\mathrm{tr}\left[ \bar{A}_\mu (-k,s')\bar{A}_\nu(k,s) \right] ~\Gamma^{\mu\nu}(k,s,s') . \label{eq:30}
\end{equation}
In this region, the propagator $G$ can be written as (See Eq.~(\ref{eq:29})):
\begin{equation}
 G(p',s;s')  =  S^{(-)}(p',s-s') + D^{(-)} (p',s+s'),
\end{equation}
and
\begin{equation}
G(p,s';s) = S^{(-)}(p,s'-s) + D^{(-)} (p,s+s'),
\end{equation}
where
\begin{equation}
 D^{(-)}(p,s+s')\equiv \mathcal{D}^{(-)}(p) ~ e^{(s'+s)\sqrt{p^2+M^2}}.
\end{equation}
Thus $\Gamma^{\mu\nu}(k,s,s')$ is
\footnote{We drop the arguments $p,p',s,s'$ for simplicity.
The symbols without prime such as $G,S^{(-)},D^{(-)}$ mean that their arguments are $(p,s;s')$.
On the other hand, the symbols with prime stand for the arguments $(p',s;s')$.}
%\mathrm{tr}\left[ \gamma^\mu G(p,s';s) \gamma^\nu G(p',s;s') \right] in Eq.~(\ref{eq:42}) for region I can be divided as follows:
\begin{eqnarray}
 &&\Gamma^{\mu\nu}(k,s,s') \\
 &=&\int \frac{d^dp}{(2\pi)^d}  \left[  \mathrm{tr}\left[ \gamma^\mu G~ \gamma^\nu G' ~\right] -\mathrm{tr}\left[ \gamma^\mu S^{(-)} \gamma^\nu S'^{(-)} \right] \right]\\
 &=& \int \frac{d^dp}{(2\pi)^d} \left[\mathrm{tr}\left[ \gamma^\mu \left(S^{(-)}+D^{(-)}\right) \gamma^\nu \left(S'^{(-)}+D'^{(-)} \right) \right] -\mathrm{tr}\left[ \gamma^\mu S^{(-)} \gamma^\nu S'^{(-)} \right] \right] \\
 &=& \int \frac{d^dp}{(2\pi)^d} \left[ \mathrm{tr}\left[ \gamma^\mu D^{(-)} \gamma^\nu D'^{(-)} \right] + \mathrm{tr}\left[ \gamma^\mu D^{(-)} \gamma^\nu S'^{(-)} \right] + \mathrm{tr}\left[ \gamma^\mu S^{(-)} \gamma^\nu D'^{(-)} \right] \right] \\
 &\equiv& \int \frac{d^dp}{(2\pi)^d} ~T_{local}^{(-)}(p,p',s,s') ,\label{eq:23}
\end{eqnarray}
where $T_{local}^{(-)}(p,p',s,s')$ depends on $s,s'$ as follows :
\begin{eqnarray}
 T_{local}^{(-)}(p,p',s,s') &=& \alpha(p,p')~ e^{(s+s')(\sqrt{p^2+M^2}+\sqrt{p'^2+M^2})} \nonumber \\
 && +~ \beta(p,p') ~ e^{(s'+s)\sqrt{p^2+M^2}} ~e^{(s'-s)\sqrt{p'^2+M^2}} \nonumber \\
 && +~ \gamma(p,p') ~e^{(s'+s)\sqrt{p'^2+M^2}}~e^{(s'-s)\sqrt{p^2+M^2}}. \label{eq:28}
\end{eqnarray}
Here, $\alpha,\beta,\gamma$ are functions of $p,p'$ and obtained from $\mathrm{tr}\left[ \gamma^\mu D^{(-)} \gamma^\nu D'^{(-)} \right]$, $\mathrm{tr}\left[ \gamma^\mu D^{(-)} \gamma^\nu S'^{(-)} \right]$, $\mathrm{tr}\left[ \gamma^\mu S^{(-)} \gamma^\nu D'^{(-)} \right]$, respectively (See Appendix.~\ref{sec:details-2-point}).
Note that the bulk term from the domain-wall fermion, $\mathrm{tr}\left[ \gamma^\mu S^{(-)} \gamma^\nu S^{(-)} \right]$, has been cancelled by the subtracting field, and there remains only $T_{local}^{(-)}$~, which damps exponentially in $s,s'$.
%We call this part ``localized part''.
%$S^{(-)}$ is the propagator with constant mass $-M$ in $2n$+1 dimensions, $\mathrm{tr}\left[ \gamma^\mu S^{(-)} \gamma^\nu S^{(-)} \right]$ in Eq.~(\ref{eq:23}) is cancelled by the subtracting field.
%Therefore, $\Gamma^{\mu\nu}(k,s,s')$ is rewritten by
%\begin{eqnarray}
% &&\left. \Gamma^{\mu\nu}(k,s,s') \right|_{s'<s<0} \nonumber \\
% &=& \int\frac{d^dp}{(2\pi)^d} \left. \left[ \mathrm{tr}\left[ \gamma^\mu G(p,s';s) \gamma^\nu G(p',s;s') \right] - (\mathrm{subtracting~field}) \right]\right|_{s'<s<0} \\ 
% &=&  \int\frac{d^dp}{(2\pi)^d} ~ T_{local}^{(-)}(p,p',s,s') .
%\end{eqnarray}
Therefore, $\Gamma^{\mu\nu}(k,s,s')$ in the region I has values only in $- 1/M \lesssim s',s < 0$ .
%
%where,
%\begin{eqnarray}
% &&\Gamma^{(\mathrm{I})\mu\nu}(k,s,s') \nonumber \\
% &=& \int\frac{d^dp}{(2\pi)^d}\left[ \mathrm{tr}\left( \gamma^\mu G^{(2)}(p,s';s) \gamma^\nu G^{(1)}(p',s;s') \right) - (\mathrm{subtracting~field}) \right] \nonumber \\ 
% &=&  \int\frac{d^dp}{(2\pi)^d}\left[ \mathrm{tr}\left( \gamma^\mu S^{(+;\leftarrow)}(p,s';s) \gamma^\nu S^{(+;\rightarrow)}(p',s;s') \right) + \mathrm{tr}\left( \gamma^\mu S^{(+;\leftarrow)}(p,s';s) \gamma^\nu D^{(+)}(p',s;s') \right) \right. \nonumber \\
% && \hspace{4em}+  \mathrm{tr}\left( \gamma^\mu D^{(+)}(p,s';s) \gamma^\nu S^{(+;\rightarrow)}(p',s;s') \right) + \mathrm{tr}\left( \gamma^\mu D^{(+)}(p,s';s) \gamma^\nu D^{(+)}(p',s;s')\right) \nonumber \\
% && \hspace{4em} \bigl. - (\mathrm{subtracting~field}) \bigr] \nonumber
%\end{eqnarray}
%with $p'=p-k$.
%%
%Here, look at $s,s'$ dependence of the each terms in the last line. %$\Gamma^{(\mathrm{I})\mu\nu}$.
%The second, third, and fourth terms have a factor ``$\exp[-(s+s')\sqrt{p_*^2+M^2}]$'' via $D^{(+)}$ ($p_*$ denotes $p$ or $p'$ ), and are not negligible only for $0<s,s'\lesssim 1/M$.
%Thus these terms are localized on domain wall $s=s'=0$.
%While, the fist term has only ``$\exp[-(s-s')\sqrt{p_*^2+M^2}]$'' via $S^{(+;\rightarrow)}$ and $S^{(+;\leftarrow)}$.
%This is the non-localized heavy fermion loop and will be cancelled by subtracting field since $S$ is precisely the $2n$-dimensional propagator with mass $M$.
%%

Using this fact, we evaluate the integral with respect to $s,s'$ in Eq.~(\ref{eq:30}) as follows.
%$\Gamma^{(\mathrm{I})\mu\nu}$ is localized on wall $s=s'=0$ and we have % in the limit of $M\rightarrow\infty$, is non-zero only at $s=s'=0$.
First, we approximate that $\bar{A}(x,s)$, which evolves by the gradient flow, is constant in the region $- 1/M \lesssim s',s < 0$.
 Thus we can write
\begin{eqnarray}
 \mathrm{I}&=& \iint_{(\mathrm{I})}  ~\mathrm{tr}\left[ \bar{A}_\mu (-k,s')\bar{A}_\nu(k,s) \right]  ~\int \frac{d^dp}{(2\pi)^d}~T_{local}^{(-)}(p,p',s,s') \\
 &\sim& \mathrm{tr}\left[ \bar{A}_\mu (-k,0)\bar{A}_\nu(k,0) \right]~ \iint _{(\mathrm{I})} ~\int \frac{d^dp}{(2\pi)^d}~T_{local}^{(-)}(p,p',s,s') \\
 &=& \mathrm{tr}\left[ A_\mu (-k)A_\nu(k) \right]~ \int \frac{d^dp}{(2\pi)^d} \iint _{(\mathrm{I})} ~T_{local}^{(-)}(p,p',s,s').
\end{eqnarray}
 The approximation ``$\sim$'' will be exact if we take the limit $M'\rightarrow\infty$.
Then, for example, for the first term in Eq.~(\ref{eq:28}), we have
%since the localized part of $\Gamma^{(\mu\nu}(k,s,s')$ have $s,s'$ dependence only as exponentials, we can pick up only the value near $s=s'=0$.
\begin{eqnarray}
&& \iint _{(\mathrm{I})}  ~ \alpha(p,p')~ e^{(s+s')(\sqrt{p^2+M^2}+\sqrt{p'^2+M^2})} \\
 &=&  \int^0_{-\infty} ds \int ^{s}_{-\infty} ds' ~\alpha(p,p')~ e^{(s+s')(\sqrt{p^2+M^2}+\sqrt{p'^2+M^2})} \label{eq:12} \\
   &=& \left. \alpha(p,p') \int^0_{-\infty} ds \left( \frac{ e^{(s'+s)(\sqrt{p^2+M^2}+\sqrt{p'^2+M^2})} }{\sqrt{p^2+M^2}+\sqrt{p'^2+M^2}}\right|_{s'=s} \right) \label{eq:33} \\
 &=& \left. \alpha(p,p') \frac{e^{2s(\sqrt{p^2+M^2}+\sqrt{p'^2+M^2})}}{2(\sqrt{p^2+M^2}+\sqrt{p'^2+M^2})^2}  \right|_{s=0} \label{eq:34} \\
 &=& \alpha(p,p')\frac{1}{2(\sqrt{p^2+M^2}+\sqrt{p'^2+M^2})^2} .
\end{eqnarray}
  The other exponentials in Eq.~(\ref{eq:28}) can be integrated in the same way and we obtain the expression Eq.~(\ref{eq:73}).
    
We denote the contribution from the region II by II:
\begin{equation}
\mathrm{II}\equiv \iint _{(\mathrm{II})} ~\mathrm{tr}\left[ \bar{A}_\mu (-k,s')\bar{A}_\nu(k,s) \right] ~\Gamma^{\mu\nu}(k,s,s').\label{eq:20}
\end{equation}
The propagator $G$ in this region is given by (See Eq.~(\ref{eq:29}))
\begin{eqnarray}
 G(p',s;s') &=& \mathcal{D}^{(-+)}(p') ~ e^{(s-s')\sqrt{p'^2+M^2}} \\
 &\equiv& D^{(-+)}(p',s-s'),
\end{eqnarray}
\begin{eqnarray}
 G(p,s';s) &=& 	\mathcal{D}^{(+-)}(p) ~ e^{(s-s')\sqrt{p^2+M^2}} \\
 &\equiv& D^{(+-)}(p,s-s').
\end{eqnarray}
Thus $\Gamma^{\mu\nu}(k,s,s')$ is
%\mathrm{tr}\left[ \gamma^\mu G(p,s';s) \gamma^\nu G(p',s;s') \right] in Eq.~(\ref{eq:42}) for region I can be divided as follows:
\begin{eqnarray}
 &&\Gamma^{\mu\nu}(k,s,s') \\
 &=&\int \frac{d^dp}{(2\pi)^d}  \left[  \mathrm{tr}\left[ \gamma^\mu G(p,s';s)~ \gamma^\nu G'(p',s;s') ~\right] -\mathrm{tr}\left[ \gamma^\mu S^{(-)} \gamma^\nu S'^{(-)} \right] \right] \\
 &=& \int \frac{d^dp}{(2\pi)^d} \left[\mathrm{tr}\left[ \gamma^\mu D^{(+-)} \gamma^\nu D'^{(-+)}  \right] -\mathrm{tr}\left[ \gamma^\mu S^{(-)} \gamma^\nu S'^{(-)} \right] \right] . \label{eq:59}
% &=& \int \frac{d^dp}{(2\pi)^d} \left[ \mathrm{tr}\left[ \gamma^\mu D^{(-)} \gamma^\nu D^{(-)} \right] + \mathrm{tr}\left[ \gamma^\mu D^{(-)} \gamma^\nu S^{(-)} \right] + \mathrm{tr}\left[ \gamma^\mu S^{(-)} \gamma^\nu D^{(-)} \right] \right] \\
% &\equiv& \int \frac{d^dp}{(2\pi)^d} ~T_{local}^{(-)}(p,p',s,s') 
\end{eqnarray}
Note that $\Gamma^{\mu\nu}$ has values only when $-1/M \lesssim s<0<s'\lesssim 1/M$ because the integrand in Eq.~(\ref{eq:59}) is proportional to $\exp[(s-s')(\sqrt{p^2+M^2}+\sqrt{p'^2+M^2})]$.
Therefore, the calculation can be performed similarly to the region I, and we obtain the resulting expression Eq.~(\ref{eq:19}).

We denote the contribution from the region III by III:
\begin{equation}
\mathrm{III}\equiv \iint_{(\mathrm{III})} ~\mathrm{tr}\left[ \bar{A}_\mu (-k,s')\bar{A}_\nu(k,s) \right] ~\Gamma^{\mu\nu}(k,s,s').\label{eq:32}
\end{equation}
In this region, the propagator $G$ can be written as 
\begin{equation}
 G(p',s;s')  = S^{(+)}(p',s-s') + D^{(+)} (p',s+s'),
\end{equation}
and
\begin{equation}
 G(p,s';s) =  S^{(+)}(p,s'-s) + D^{(+)} (p,s+s') ,
\end{equation}
where
\begin{equation}
D^{(+)}(p,s+s') \equiv \mathcal{D}^{(+)}(p) ~ e^{-(s'+s)\sqrt{p^2+M^2}}.
\end{equation}
Thus $\Gamma^{\mu\nu}(k,s,s')$ is calculated as
%\mathrm{tr}\left[ \gamma^\mu G(p,s';s) \gamma^\nu G(p',s;s') \right] in Eq.~(\ref{eq:42}) for region I can be divided as follows:
\begin{eqnarray}
 &&\Gamma^{\mu\nu}(k,s,s') \\
 &=&\int \frac{d^dp}{(2\pi)^d}  \left[  \mathrm{tr}\left[ \gamma^\mu G~ \gamma^\nu G' ~\right] -\mathrm{tr}\left[ \gamma^\mu S^{(-)} \gamma^\nu S'^{(-)} \right] \right]\\
 &=& \int \frac{d^dp}{(2\pi)^d} \left[\mathrm{tr}\left[ \gamma^\mu \left(S^{(+)}+D^{(+)}\right) \gamma^\nu \left(S'^{(+)}+D'^{(+)} \right) \right] -\mathrm{tr}\left[ \gamma^\mu S^{(-)} \gamma^\nu S'^{(-)} \right] \right] \\
 &=& \int \frac{d^dp}{(2\pi)^d} \left[ \mathrm{tr}\left[ \gamma^\mu D^{(+)} \gamma^\nu D'^{(+)} \right] + \mathrm{tr}\left[ \gamma^\mu D^{(+)} \gamma^\nu S'^{(+)} \right] + \mathrm{tr}\left[ \gamma^\mu S^{(+)} \gamma^\nu D'^{(+)} \right] \right] \nonumber \\
 && + \int \frac{d^dp}{(2\pi)^d} \left[  \mathrm{tr}\left[ \gamma^\mu S^{(+)} \gamma^\nu S'^{(+)} \right]  -\mathrm{tr}\left[ \gamma^\mu S^{(-)} \gamma^\nu S'^{(-)} \right] \right] \\
 &\equiv&  \int \frac{d^dp}{(2\pi)^d} \left[ T_{local}^{(+)} (p,p',s,s')+ T_{bulk}(p,p',s,s')\right] , \label{eq:4}
\end{eqnarray}
where
\begin{equation}
 T_{bulk} \equiv \mathrm{tr}\left[ \gamma^\mu S^{(+)} \gamma^\nu S'^{(+)} \right]  -\mathrm{tr}\left[ \gamma^\mu S^{(-)} \gamma^\nu S'^{(-)} \right] , \label{eq:51}
\end{equation}
\begin{eqnarray}
 T_{local}^{(+)} &\equiv& \mathrm{tr}\left[ \gamma^\mu D^{(+)} \gamma^\nu D'^{(+)} \right] + \mathrm{tr}\left[ \gamma^\mu D^{(+)} \gamma^\nu S'^{(+)} \right] + \mathrm{tr}\left[ \gamma^\mu S^{(+)} \gamma^\nu D'^{(+)} \right] \\
 &=& \alpha(p,p')~ e^{-(s+s')(\sqrt{p^2+M^2}+\sqrt{p'^2+M^2})} \nonumber \\
 && +~ \beta(p,p') ~ e^{-(s'+s)\sqrt{p^2+M^2}} ~e^{(s'-s)\sqrt{p'^2+M^2}} \nonumber \\
 && +~ \gamma(p,p') ~e^{-(s'+s)\sqrt{p'^2+M^2}}~e^{(s'-s)\sqrt{p^2+M^2}} . \label{eq:5}
\end{eqnarray}
Here, $\alpha,\beta,\gamma$ in Eq.~(\ref{eq:5}) are the same as those in Eq.~(\ref{eq:28})(See Appendix.~\ref{sec:details-2-point}).
 Note that $T_{local}^{(+)}$ , which is localized on $s=0$, gives the same contribution as $T_{local}^{(-)}$ after integrating over $s$ and $s'$.
On the other hand, $T_{bulk}$ is the bulk term, which does not vanish unlike the region I because of the opposite signs of the masses.
We will discuss this point in the next subsection.

Next, we consider the other regions $\mathrm{I}',\mathrm{II}',\mathrm{III}'$.
The net effect of interchanging $s\leftrightarrow s'$ is to change the signs of $\gamma^5$ in $S^{(+)}$ and $S^{(-)}$ (See Eq.~(\ref{eq:10})).
%The latter change does not affect the resulting expressions after integrating with respect to $s,s'$.
Therefore the contributions from $\mathrm{I}',\mathrm{II}',\mathrm{III}'$ are obtained from those of $\mathrm{I},\mathrm{II},\mathrm{III}$, respectively, by changing the signs of $\gamma^5$ in $S^{(+)}$ and $S^{(-)}$. (See the discussions below Eq.~(\ref{eq:73}) and Eq.~(\ref{eq:19}))

All contributions (Eq.~(\ref{eq:73}),~(\ref{eq:19}),~(\ref{eq:4})) sum up to
\begin{eqnarray}
I_2 &=& \int\frac{d^dk}{(2\pi)^d} \left( \mathrm{I}+\mathrm{II}+\mathrm{III} + \mathrm{I}' + \mathrm{II}' + \mathrm{III}' \right) \\
 &=& \int \frac{d^dk}{(2\pi)^d}~\mathrm{tr} [A_\mu(-k)A_\nu(k)] \left(\Pi^{\mu\nu}_{{(non-anomalous)}}+ \Pi^{\mu\nu}_{{(anomalous)}}\right) + I_2^{bulk}(\bar{A}),\label{eq:55}
\end{eqnarray}
where 
\begin{equation}
 \Pi^{\mu\nu}_{{(non-anomalous)}} \equiv \int \frac{d^dp}{(2\pi)^d} \left[ \frac{2^n M^2(p^2+p'^2+M^2)N^{\mu\nu}}{2p^2p'^2 (p^2+M^2)(p'^2+M^2)}  -\frac{2^n M^2 \delta^{\mu\nu}}{2(p^2+M^2)(p'^2+M^2)} \right] \label{eq:47},
\end{equation}
%%%%
\begin{equation} 
 \Pi^{\mu\nu}_{{(anomalous)}} \equiv \int \frac{d^dp}{(2\pi)^d} \frac{-M~\mathrm{tr}\left[\gamma^\mu\slashed{p}\gamma^\nu\slashed{p}'\gamma^5 \right]\left(p^2+p'^2+M^2+\sqrt{p^2+M^2}\sqrt{p'^2+M^2}\right)}{2p^2p'^2\sqrt{p^2+M^2}\sqrt{p'^2+M^2}(\sqrt{p^2+M^2}+\sqrt{p'^2+M^2})} \label{eq:25},
\end{equation}
\begin{equation}
I_2^{bulk} (\bar{A})\equiv \int \frac{d^dk}{(2\pi)^d} \int_0^\infty ds \int_0^\infty ds' ~\mathrm{tr} [\bar{A}_\mu(-k,s')\bar{A}_\nu(k,s)]\int\frac{d^dp}{(2\pi)^d} ~ T_{bulk}(p,p',s,s').\label{eq:57}
\end{equation}
Here, $N^{\mu\nu} \equiv p\cdot p'\delta^{\mu\nu}-p^\mu p'^\nu -p^\nu p'^\mu $.
The first term in Eq.~(\ref{eq:55}) represents the contribution from the localized terms.
$\Pi^{\mu\nu}_{{(anomalous)}}$ and $\Pi^{\mu\nu}_{{(non-anomalous)}}$ are the parts with and without $\gamma^5$, respectively.
As we will see in the next subsection,
\begin{equation}
  \Pi^{\mu\nu}_{{(non-anomalous)}}+ \Pi^{\mu\nu}_{{(anomalous)}} \label{eq:60}
\end{equation}
is equal to the vacuum polarization of a left-handed chiral fermion with a Pauli-Villars-like regulator of mass $M$.
$I_2^{bulk}$ represents the contribution from the bulk region $0<s,s'<\infty$.

Note that there are no leading UV divergences, terms that have degree of divergence $d-2$, in Eq.~(\ref{eq:47})-(\ref{eq:57}).
Therefore, all UV divergences are cancelled by the Pauli-Villars pairs under the conditions, Eq.~(\ref{eq:50}).
 %and thus we regard these integrals as UV finite.
%Indeed the leading divergences, which are independent of $M$, are cancelled.
%
%looks complicated, but is equal to an anomalous part of the 1-loop 2-point function taking the limit $M\rightarrow\infty$.
%We prove this in the next subsection.

\subsection{Comparison with chiral fermion in $2n$ dimensions} \label{sec:comp-with-chir}
\quad
%$\Pi^{\mu\nu}_{{(non-anomalous)}}$ is precisely equal to a non-anomalous (i,e, not including $\gamma^5$) part of the 1-loop 2-point function of the chiral fermion with a Pauli-Villars field of mass $M$.
%We discuss $\Pi^{\mu\nu}_{{(anomalous)}}$ in the next subsection.
%
We first consider the vacuum polarization of a left-handed chiral fermion:
\begin{equation}
\int \frac{d^dp}{(2\pi)^d}  \mathrm{tr} \left[\gamma^\mu P_- ~\frac{i\slashed{p}}{p^2} ~\gamma^\nu P_- ~\frac{i\slashed{p}'}{p'^2}\right] \label{eq:58} ,
\end{equation}
where $P_- = \frac{1-\gamma^5}{2}$ .
By introducing one Pauli-Villars field, we have
\begin{eqnarray}
 && \int \frac{d^dp}{(2\pi)^d} \left( \mathrm{tr} \left[\gamma^\mu P_- ~\frac{i\slashed{p}}{p^2} ~\gamma^\nu P_- ~\frac{i\slashed{p}'}{p'^2}\right] -  \mathrm{tr} \left[\gamma^\mu P_- ~\frac{i\slashed{p}+M}{p^2+M^2} \gamma^\nu P_-~ \frac{i\slashed{p}'+ M }{p'^2+M^2}\right] \right) \\
 &&\nonumber \\
 &=& V^{\mu\nu}_{(non-anomalous)} + V^{\mu\nu}_{(anomalous)}\label{eq:64},
\end{eqnarray}
where
\begin{equation}
 V^{\mu\nu}_{(non-anomalous)} \equiv \int \frac{d^dp}{(2\pi)^d} \left( \frac{- M^2(p^2+p'^2+M^2)~\mathrm{tr}\left[\gamma^\mu \slashed{p}\gamma^\nu\slashed{p}'\right]}{2p^2p'^2(p^2+M^2)(p'^2 +M^2)} - \frac{ M^2 \mathrm{tr}\left[\gamma^\mu \gamma^\nu\right]}{2(p^2+M^2)(p'^2 +M^2)}  \right) \label{eq:65} ,
\end{equation}
\begin{equation}
 V^{\mu\nu}_{(anomalous)} \equiv \int \frac{d^dp}{(2\pi)^d} \left( -\frac{\mathrm{tr}\left[\gamma^\mu\slashed{p}\gamma^\nu\slashed{p}'\gamma^5\right]}{2p^2p'^2} +\frac{\mathrm{tr}\left[\gamma^\mu\slashed{p}\gamma^\nu\slashed{p}'\gamma^5\right]}{2(p^2+M^2)(p'^2+M^2)} \right)\label{eq:37}.
\end{equation}
The non-anomalous part $V^{\mu\nu}_{(non-anomalous)}$ is precisely equal to Eq.~(\ref{eq:47}).
%Thus we consider an anomalous part of Eq.~(\ref{eq:37}):
%\begin{eqnarray}
% && \int \frac{d^dp}{(2\pi)^d} \left( -\frac{\mathrm{tr}\left[\gamma^\mu\slashed{p}\gamma^\nu\slashed{p}'\gamma^5\right]}{2p^2p'^2} +\frac{\mathrm{tr}\left[\gamma^\mu\slashed{p}\gamma^\nu\slashed{p}'\gamma^5\right]}{2(p^2+M^2)(p'^2+M^2)} \right) \\
% &=& - \int \frac{d^dp}{(2\pi)^d} \frac{M^2(p^2+p'^2+M^2)~\mathrm{tr}\left[\gamma^\mu\slashed{p}\gamma^\nu\slashed{p}'\gamma^5\right]}{2p^2p'^2(p^2+M^2)(p'^2+M^2)} \label{eq:38} .
%\end{eqnarray}

We evaluate the difference of Eq.~(\ref{eq:25}) and Eq.~(\ref{eq:37}), and show it is zero in the limit of $M\rightarrow\infty$.
This is trivial for $d>2$ since both of them vanish.
Thus we consider the case $d=2$.
The difference is calculated as:
\begin{eqnarray}
 F_M(k) &\equiv& V^{\mu\nu}_{(anomalous)} - \Pi^{\mu\nu}_{{(anomalous)}} \nonumber \\
 &=&\int \frac{d^2p}{(2\pi)^2} \left[- \frac{M^2(p^2+p'^2+M^2)~\mathrm{tr}\left[\gamma^\mu\slashed{p}\gamma^\nu\slashed{p}'\gamma^5\right]}{2p^2p'^2(p^2+M^2)(p'^2+M^2)} \right. \nonumber \\
 &&\left. + \frac{M~\mathrm{tr}\left[\gamma^\mu\slashed{p}\gamma^\nu\slashed{p}'\gamma^5 \right]\left(p^2+p'^2+M^2+\sqrt{p^2+M^2}\sqrt{p'^2+M^2}\right)}{2p^2p'^2\sqrt{p^2+M^2}\sqrt{p'^2+M^2}(\sqrt{p^2+M^2}+\sqrt{p'^2+M^2})}  \right] \nonumber .\\
\end{eqnarray}
 Because this integral is finite at $k=0$, we can expand it around $k=0$:
 \begin{equation}
  F_M(k) = F_M(0) + \mathcal{O}\left(\frac{k}{M}\right),
 \end{equation}
 where,
 \begin{eqnarray}
 && F_M(0) \nonumber \\
  &=& \int \frac{d^2p}{(2\pi)^2} \frac{M~\mathrm{tr}\left[\gamma^\mu\slashed{p}\gamma^\nu\slashed{p}\gamma^5\right]}{(p^2)^2(p^2+M^2)^2}\left( - \frac{1}{2}M(2p^2+M^2) + \frac{1}{4}\sqrt{p^2+M^2}(3p^2+2M^2) \right) \\
  &\propto& \mathrm{tr} \left[\gamma^\mu\gamma^\lambda\gamma^\nu\gamma_\lambda\gamma^5\right] \\
  &=& 0 .
 \end{eqnarray}
 Thus we obtain
\begin{equation}
 \lim_{M \to \infty}F_M(k) = 0 .
\end{equation}
Therefore, Eq.~(\ref{eq:60}) is equal to Eq.~(\ref{eq:64}) in the limit of $M\rightarrow \infty$.
%obtain the result that the contribution to $I_2$ from the domain wall $s=0$ is equal to the vacuum polarization loop of the left-handed chiral fermion with a Pauli-Villars field.

%we consider $I_m$ for general cases ($m>2$).
We extend the above result to general cases $m>2$.
It is expected that $I_m$ given by Eq.~(\ref{eq:56}) is also written as
%We extend this result to the general cases $m>2$, and thus can write
\begin{equation}
 I_m = I_m^{s=0} + I_m^{bulk} \label{eq:66},
\end{equation}
where $I_m^{s=0}$ is the $m$-vertex loop of the left-handed chiral fermion with the Pauli-Villars field on the domain wall $s=0$.
$I_m^{bulk}$ is the $m$-vertex loop of the heavy mode and the subtracting field in the bulk region $0<s<\infty$.
Note that the divergent terms that are included in $I_m^{s=0}$ and $I_m^{bulk}$ will be cancelled by adding the Pauli-Villars pairs.
%\begin{eqnarray}
% I_m^{bulk} &=&   \left( \prod_i^m \int \frac{d^dk_i}{(2\pi)^d} \right) \left( \prod_i^m \int_{0}^\infty ds_i \right) (2\pi)^d~\delta^{(d)}(\sum_i k_i) \nonumber \\
% &&\hspace{2em}\times \mathrm{tr} \left[\bar{A}_{\mu_1}(k_1,s_1)\cdots \bar{A}_{\mu_m}(k_m,s_m)\right]~ \nonumber \\
% && \hspace{2em} \times\left( \mathrm{tr}\left[\gamma^{\mu_1} S^{(+)} \cdots \gamma^{\mu_m} S^{(+)}\right]  -   \mathrm{tr} \left[ \gamma^{\mu_1} S^{(-)} \cdots \gamma^{\mu_m} S^{(-)} \right] \right) \label{eq:62} .
 %\end{eqnarray}
 %In order to obtain the effective action Eq.~(\ref{eq:68}) in the limit of $L\rightarrow\infty$,
Then, from Eq.~(\ref{eq:66}), we have
 \begin{eqnarray}
  && \lim_{L \to \infty} \left[\mathrm{Tr} \log (\slashed{D}_{2n+1} -\epsilon(s)M) - \mathrm{Tr} \log (\slashed{D}_{2n+1} + M )\right]\\
  &=& \sum_m \frac{1}{m} I_m^{s=0} + \sum_m \frac{1}{m} I_m^{bulk} \label{eq:70}.
 \end{eqnarray}
 %For the contribution from the wall $s=0$,
 The first term in Eq.~(\ref{eq:70}) can be regarded as
\begin{equation}
 \sum_m \frac{1}{m} I_m^{s=0} = \mathrm{Tr}\log (\slashed{D}_{2n}P_- + \slashed{\partial}_{2n}P_+ ) - \mathrm{Tr}\log (\slashed{D}_{2n}P_- + \slashed{\partial}_{2n}P_+  -M) .\label{eq:69}
\end{equation}
Here, the first term and the second term in Eq.~(\ref{eq:69}) are the effective actions of the left-handed chiral fermion and the Pauli-Villars field, respectively.
%
%\begin{eqnarray}
% \sum \frac{1}{m} I_m &=& \sum_m \frac{1}{m} I_m^{s=0} + \sum_m \frac{1}{m} I_m^{bulk} \\
% &=& \mathrm{Tr}\log (\slashed{D}_{2n}P_- + \slashed{\partial}_{2n}P_+ ) - \mathrm{Tr}\log (\slashed{D}_{2n}P_- + \slashed{\partial}_{2n}P_+  -M) \nonumber \\
% &&+ \sum_m \frac{1}{m} I_m^{bulk} ,
%\end{eqnarray}
%We can expect the similar results for the general $I_m$ and obtain
%\begin{equation}
% \sum _m \frac{1}{m} I_m = \mathrm{Tr} \log \left(\slashed{D}_{2n}P_- + \slashed{\partial}_{2n}P_+ \right) + \sum_m \frac{1}{m} S_m^{bulk} . \label{eq:61}
%\end{equation}
On the other hand, the second term in Eq.~(\ref{eq:70}) can be written as\cite{Callan:1984sa,Deser:1981wh}:
\begin{equation}
 \sum_m \frac{1}{m} I_m^{bulk} = S_{2n+1}^{(CS)} + \delta S_{2n+1}(M),
\end{equation}
where $S_{2n+1}^{(CS)}$ is the Chern-Simons term given by Eq.~(\ref{eq:35}).
The ultra-violet (UV) divergence in $\delta S_{2n+1}(M)$ is cancelled after combining with the Pauli-Villars pairs.
%the sum of the bulk contribution $\sum_m \frac{1}{m} I_m^{bulk} $ is equal to the Chern-Simons term $S^{(CS)}_{2n+1}$ (Eq.~(\ref{eq:35})).
%Therefore, we obtain
%\begin{eqnarray}
% \sum \frac{1}{m} I_m  &=& \mathrm{Tr}\log (\slashed{D}_{2n}P_- + \slashed{\partial}_{2n}P_+ ) - \mathrm{Tr}\log (\slashed{D}_{2n}P_- + \slashed{\partial}_{2n}P_+  -M) \nonumber \\
% &&+ S^{(CS)}_{2n+1} .
%\end{eqnarray}

So far, we have neglected the domain wall $s=L$ by taking the limit $L\rightarrow\infty$.
There, the similar result for the right-handed chiral fermion to Eq.~(\ref{eq:69}) should be obtained.
Therefore the effective action Eq.~(\ref{eq:68}) is 
%One is equal to the effective action of the left-handed chiral fermion with the Pauli-Villars field on $s=0$.
%The other is the Chern-Simons term in the bulk $0<s<L$.
%At $s=L$, the same result for the right-handed chiral fermion will be obtained.
\begin{eqnarray}
 && \mathrm{Tr}\log (\slashed{D}_{2n+1}-\epsilon(s)M) - \mathrm{Tr}\log (\slashed{D}_{2n+1} +M) \\
 &=&  \mathrm{Tr}\log (\slashed{D}_{2n}P_- + \slashed{\partial}_{2n}P_+ )_{s=0} - \mathrm{Tr}\log (\slashed{D}_{2n}P_- + \slashed{\partial}_{2n}P_+ -M)_{s=0} \nonumber \\
 &&+ \mathrm{Tr}\log (\slashed{D}_{2n}P_+  + \slashed{\partial}_{2n}P_-)_{s=L} - \mathrm{Tr}\log (\slashed{D}_{2n}P_+ + \slashed{\partial}_{2n}P_- -M)_{s=L} \nonumber \\
 &&+ S^{(CS)}_{2n+1} +  \delta S_{2n+1}(M).\label{eq:44}
\end{eqnarray}
Here, $()_{s=0}$ and $()_{s=L}$ stand for substituting the gauge field $\bar{A}(x,s=0)$ and $\bar{A}(x,s=L)$ into the covariant derivative, respectively.
%$P_-$ and $P_+$ denote chiral projection operators.
%Therefore, we have
%\begin{equation}
% \sum _m \frac{1}{m} I_m = \mathrm{Tr} \log \left(\slashed{D}_{2n}P_- + \slashed{\partial}_{2n}P_+ \right) + S_{2n+1}^{(CS)} . \label{eq:63}
%\end{equation}
%
%In this section, we calculate the effective action only to the quadratic order.
%Assuming this result is correct to all orders, we obtain
%Manifestly, Pauli-Villars fields in Eq.~(\ref{eq:44}) have gauge variant coupling, but the variance is cancelled by the Chern-Simons term as argued in Ref.\cite{Callan:1984sa}.
%%

By adding the Pauli-Villars pairs, the regularized effective action is obtained as follows.
For the sake of simplicity, we write Eq.~(\ref{fermion_det_reg5}) as
\begin{equation}
 \log \Delta (A)_\mathrm{reg.} =  \sum_{i=0} C_i \left[ \mathrm{Tr}\log\left( \slashed{D}_{2n+1} - M_i \epsilon(s)\right)- \mathrm{Tr}\log\left( \slashed{D}_{2n+1} + M_i \right) \right] ,\label{eq:71}
\end{equation}
where $C_0 = 1$ and $ M_0=M$.
By applying Eq.~(\ref{eq:44}) to each pair, we have
\begin{eqnarray}
\log \Delta (A)_\mathrm{reg.}  &=&  \sum_{i=0} C_i \left[ \mathrm{Tr}\log (\slashed{D}_{2n}P_- + \slashed{\partial}_{2n}P_+ )_{s=0} - \mathrm{Tr}\log (\slashed{D}_{2n}P_- + \slashed{\partial}_{2n}P_+  -M_i)_{s=0} \right] \nonumber \\
 && + \sum_{i=0} C_i \left[ \mathrm{Tr}\log (\slashed{D}_{2n}P_+  + \slashed{\partial}_{2n}P_-)_{s=L} - \mathrm{Tr}\log (\slashed{D}_{2n}P_+ + \slashed{\partial}_{2n}P_-  -M_i)_{s=L} \right] \nonumber \\
 && +\sum_{i=0} C_i ~S^{(CS)}_{2n+1} + \sum_{i=0} C_i ~\delta S_{2n+1}(M_i) \label{eq:21} \\
 &=& \mathrm{Tr}\log (\slashed{D}_{2n}P_- + \slashed{\partial}_{2n}P_+ )_{s=0} - \sum_{i=0} C_i ~ \mathrm{Tr}\log (\slashed{D}_{2n}P_-  + \slashed{\partial}_{2n}P_+ -M_i)_{s=0}  \nonumber \\
 && + \mathrm{Tr}\log (\slashed{D}_{2n}P_+ + \slashed{\partial}_{2n}P_-)_{s=L} - \sum_{i=0} C_i ~ \mathrm{Tr}\log (\slashed{D}_{2n}P_+  + \slashed{\partial}_{2n}P_- -M_i)_{s=L} \nonumber \\
 &&+ S^{(CS)}_{2n+1} \label{eq:22} ~.
\end{eqnarray}
The last term in Eq.~(\ref{eq:21}) is UV-finite and vanishes in the limit of $M,M_i \rightarrow\infty$, which we drop in the following expressions.
As argued in Sec.\ref{sec:regul-form}, the extra massless modes and Chern-Simons terms have vanished by the condition $\sum_{i=1} C_i=0$.
Thus there are no artificial degrees of freedom.
In addition, the regularized effective action Eq.~(\ref{eq:22}) converges under the condition, Eq.~(\ref{eq:50}).

Note that Eq.~(\ref{eq:22}) is gauge invariant because gauge anomalies from the three lines are cancelled.
For example, for $n=2$, the gauge variation of the Chern-Simons term is
\begin{equation}
 \delta_\chi S_{5}^{(CS)} = \frac{-1}{48\pi^2} \int d^4x~\epsilon_{\mu\nu\lambda\rho}~ \left. \mathrm{tr}\left[\chi ~\partial_\mu  \left(\bar{A}_\nu\bar{A}_\lambda\bar{A}_\rho + 2\bar{A}_\nu\partial_\lambda\bar{A}_\rho\right) \right]\right|_{s=0}^{s=L},\label{eq:105}
\end{equation}
where $\chi$ is the gauge function.
On the other hand, the first and second lines in Eq.~(\ref{eq:22}) give the anomaly of the left- and right-handed chiral fermions in 4 dimensions, respectively, which cancel with Eq.~(\ref{eq:105}).
%This cancels with the contributions from the first and second lines in Eq.~(\ref{eq:22}), which are the anomalies of the left- and right-handed chiral fermions in 4 dimensions, respectively.
 %as argued in Ref.\cite{Grabowska:2015qpk}.
This cancellation agrees with the manifestly gauge invariant construction, Eq.~(\ref{fermion_det_reg5}).
%We have imposed
%\begin{eqnarray}
% \begin{split}
% M + \sum_i C_i M_i &=& 0  \\
% M^2+\sum_i C_i (M_i)^2 &=& 0  \\
% M^3+\sum_i C_i (M_i)^3 &=& 0  \\
%  \vdots && ,
% \end{split} \label{eq:49}
%\end{eqnarray}
%that are precisely the conditions for the divergences in Eq.~(\ref{eq:22}) to vanish.
%
%Therefore, we conclude that massless modes on the domain walls look like chiral fermions even at the quantum level and that our regularization argued in Sec.~\ref{sec:regul-form} is successful.
%%%%%
\section{Axial-vector current in vector-like gauge theory}\label{sec:constr-u1-axial}
\quad
We investigate the consistency of this formulation by introducing two sets of domain-wall fermions belonging to complex conjugate representations.
%By taking the limit $M\rightarrow\infty$, we obtain an effective theory consisting of two left-handed physical fermions and two right-handed fluff fermions on each wall.
%If the fluffs are decoupled correctly, this formulation is equivalent to the $2n$-dimensional vector-like gauge theory after we apply charge conjugation to one of the left-handed physical fermions.  
%
As a simple example, we consider a 5-dimensional U(1) gauge theory.
We assume that each set of fermions consists of a domain-wall fermion, a subtracting field and Pauli-Villars pairs.
The two domain-wall fermions $\psi$ and $\psi'$ have U(1) charge $\pm 1$, respectively.
%In the following argument, we concentrate on the physical contribution from the domain-wall fermions $\psi, \psi'$.
While left-handed physical fermions are localized on $s=0$, right-handed fluff fermions are localized on $s=L$.
We denote the former coming from $\psi$ and $\psi'$ by $q_L$ and $q'_L$, respectively.
Because the gradient flow makes the fluff fermions decouple, we obtain a 4-dimensional effective theory consisting of the left-handed physical fermions $q_L,q'_L$.
Here, the Chern-Simons term vanishes due to the representations, and the effective theory is equivalent to the vector-like theory after applying the charge conjugation: $q_R \equiv q_L'^C$.
In the following, we will show that the axial-vector current that is defined naturally does not reproduce the correct anomaly\cite{Suzuki:2016}.
%%
%%
%
%As a naive candidate of the U(1) axial-vector current,

One natural way to define such current is to introduce a fictitious U(1) gauge field $B_\mu$ that couples to $q_L$ and $q_L'$ with charge +1.
Then the current is defined by the variation with respect to the gauge field $B_\mu(x)$.
In order to realize it, we consider the bulk U(1) gauge field $\bar{B}_\mu(x,s)$ that couples to $\psi$ and $\psi'$ with charge +1.
%We define the current as follows.
%We introduce a fictitious U(1) gauge field $\bar{B}_\mu(x,s)$ that couples to both $\psi$ and $\psi'$ with charge +1.
%
We assume that $\bar{B}_\mu$ also evolves by the gradient flow from $s=0$ to $s=\pm L$:
\begin{equation}
 \partial_s \bar{B}_\nu=\frac{\epsilon(s)}{M''}\partial_\mu\bar{F}^{(B)}_{\mu\nu}~,
\end{equation}
with $\mu,\nu=1,\cdots,4$, $\bar{B}_\mu(x,0)=B_\mu(x)$ and $\bar{B}_5=0$.
$\bar{F}^{(B)}_{\mu\nu}$ denotes the field strength of $\bar{B}_\mu$, and $M'' \gg M$ as $M'$ in Eq.~(\ref{eq:91}).
%We will take the limit $M'\rightarrow\infty$ later as well as $\bar{A}_\mu$.
%Note that there is an ambiguity of the flow parameter $\xi'$, which is less than unity.
Then, we define $ J_\mu^B (x)$ by
%\begin{equation}
% \langle j _\mu^B (x,s)\rangle _{\bar{A}}  \equiv \left. \frac{\delta S_{eff}[\bar{A},\bar{B}]}{\delta \bar{B}_{\mu}(x,s)}\right|_{\bar{B}_\mu=0}
%\end{equation}
%
%
%Then, we define the 4-dimensional fermion current by the variation with respect to $B_\mu(x)$:
\begin{equation}
 \langle   J_\mu^B (x) \rangle _{A} \equiv \left. \frac{\delta S_{eff}[\bar{A},\bar{B}]}{\delta B_{\mu}(x)}\right|_{B_\mu=0}, \label{eq:88}
\end{equation}
where $\mu = 1,\cdots,4$.
$S_{eff}[\bar{A},\bar{B}]$ is the effective action obtained by integrating out $\psi$ and $\psi'$.
The symbol $\langle \rangle _{A}$ stands for the expectation value in the presence of the background gauge field $A_\mu$, which we drop in the expressions below.
$J_\mu^B$ seems to be the U(1) axial-vector current:
\begin{eqnarray}
 J_\mu^B (x) &\sim& \bar{q}_L \gamma^\mu q_L + \bar{q}'_L \gamma^\mu q'_L  \label{eq:87}\\
 &=&\bar{q}_L \gamma^\mu q_L - \bar{q}_R \gamma^\mu q_R ~. \label{eq:31}
\end{eqnarray}
%where ``$\sim$'' stands for naive equality.
%, and we have neglected the fluff fermions.
%$q_R \equiv q'^{C}_L $.
%Thus, from the viewpoint of the 4-dimensional effective theory, this looks like the U(1) axial-vector current.
However, it does not reproduce the correct axial anomaly.
%
%
%
%
%
%
%
%
%We first consider the slow limit $\xi' \rightarrow 0$, so that $\bar{B}_\mu$ is constant in the $s$ direction : $\bar{B}_\mu(x,s)= B_\mu(x)$.
Indeed, as we will see below, $J_\mu^B$ is exactly conserved\cite{Suzuki:2016}:
\begin{equation}
 \partial_\mu J_\mu^B (x) = 0 .
\end{equation}
%and can not be regarded as the axial-vector current in contrast to Eq.~(\ref{eq:31}).
%On the other hand,
On the other hand, from the viewpoint of the 5-dimensional theory, this conservation is natural because this current is a Noether current of this system.
In order to solve this paradox, we investigate the mechanism of this conservation.

First we discuss how the effective action changes under the gauge transformation of $B_\mu(x)$,
\begin{equation}
 B_\mu(x)\mapsto B_\mu(x) + \partial_\mu \chi(x) \label{eq:96} .
\end{equation}
Because $\bar{B}_\mu(x,s)$ is changed as
\begin{equation}
 \bar{B}_\mu(x,s) \mapsto \bar{B}_\mu(x,s) + \partial_\mu\chi(x) \label{eq:97},
\end{equation}
the variation of the effective action $S_{eff}[\bar{A},\bar{B}]$ can be written in the following two ways:
\begin{eqnarray}
 \delta S_{eff} &=& \int d^4x ~\partial_\mu \chi(x)~ \frac{\delta S_{eff}[\bar{A},\bar{B}]}{\delta B_\mu(x)} \\ \label{eq:98}
  &=& \int d^4x \int ds ~ \partial_\mu \chi(x) ~\frac{\delta S_{eff}[\bar{A},\bar{B}]}{\delta\bar{B}_\mu(x,s)} \label{eq:99}.
\end{eqnarray}
%We denote the gauge parameter of $\bar{B}_\mu$ by $\lambda$:
%\begin{equation}
% \bar{B}_\mu (x,s) = \bar{B}_\mu' (x,s) + \partial_\mu \lambda(x,s),\label{eq:92}
%\end{equation}
%where $\bar{B}_\mu'$ is the physical degree of freedom, and $\lambda(x,s)$ is invariant under the gradient flow : $\lambda(x,s) = \lambda(x,0) \equiv \lambda(x)$.
%From Eq.~(\ref{eq:88}), the divergence of $J_\mu^B$ is calculated by the gauge variation of $S_{eff}$:
%\begin{eqnarray}
% \frac{\delta S_{eff}[\bar{A},\bar{B}]}{\delta \lambda(x)} &=&    \int d^4y~ \frac{\delta S_{eff}[\bar{A},\bar{B}]}{\delta B_\mu(y)} \frac{\delta B_\mu(y)}{\delta \lambda(x)} \\
% &=&  \int d^4y~ J_\mu^B(y) \frac{\delta B_\mu(y)}{\delta \lambda(x)} \\
% &=& \int d^4 y ~ J_\mu^B(y) ~\partial_\mu^{(y)} \delta^4(y-x) \\
%  &=& -  \partial_\mu J_\mu^B(x) , \label{eq:93}
%\end{eqnarray}
%where we have dropped the symbol ``$|_{\bar{B} = 0}$''.
%On the other hand, from Eq.~(\ref{eq:89}),
%\begin{eqnarray}
%  \frac{\delta S_{eff}[\bar{A},\bar{B}]}{\delta \lambda(x)}  &=&   \int d^4y \int ds ~ \frac{\delta S_{eff}[\bar{A},\bar{B}]}{\delta\bar{B}_\mu(y,s)} \frac{\delta \bar{B}_\mu(y,s)}{\delta \lambda(x)} \\
% &=&   \int d^4y \int ds ~ j_\mu^B(y,s) \frac{\delta \bar{B}_\mu(y,s)}{\delta \lambda(x)} \\
%  &=& - \int ds ~ \partial_\mu ~ j_\mu^B(x,s)  \label{eq:94} .
%\end{eqnarray}
Thus we obtain
\begin{equation}
 \partial_\mu J_\mu^B(x) = \int _0^L ds ~ \partial_\mu ~ j_\mu^B(x,s) \label{eq:95},
\end{equation}
where
\begin{equation}
 j_\mu^B(x,s) \equiv \left. \frac{\delta S_{eff}[\bar{A},\bar{B}]}{\delta\bar{B}_\mu(x,s)}\right|_{\bar{B}_\mu=0} \label{eq:100} .
\end{equation}
Note that the region $-L<s<0$ has no contribution to Eq.~(\ref{eq:95}) because no terms are induced there as we have seen in Sec.~\ref{sec:culc-effect-acti}.
The above expression indicates that there is a contribution from the bulk to the divergence of the current as well as that from the domain wall, Eq.~(\ref{eq:87}).

As we have seen in the previous section, $S_{eff}[\bar{A},\bar{B}]$ consists of the effective action of the chiral fermions on $s=0,L$ and the Chern-Simons term in the bulk, in the limit of $M\rightarrow\infty$.
Thus we can write
\begin{eqnarray}
\int_0^L ds  ~j_\mu^B (x,s)  &=& J_\mu^{(q_L,q_R)}(x) + J_\mu^{(\mathrm{fluff})}(x) + J^{(CS)}_\mu (x) ,\label{eq:27}
\end{eqnarray}
where $J_\mu^{(q_L,q_R)},~ J_\mu^{(\mathrm{fluff})}$ are currents of the chiral fermions on each boundary, and
%, and
\begin{equation}
 J_\mu^{(CS)}(x) \equiv \int _0^L ds ~ j_\mu^{(CS)}(x,s) \label{eq:101}.
\end{equation}
$ j_\mu^{(CS)}(x,s)$ is the Chern-Simons current:
\begin{eqnarray}
 j_\mu^{(CS)}(x,s) &\equiv&  \left. \frac{\delta}{\delta \bar{B}_\mu} S^{(CS)}_5(\bar{A},\bar{B})  \right|_{\bar{B}=0} \label{eq:43} \\
&=& \frac{-1}{24\pi^2}~\left. \frac{\delta}{\delta \bar{B}_\mu}\int \omega_5(\bar{A},\bar{B}) \right|_{\bar{B}=0} . \label{eq:26}
\end{eqnarray}
%The factor 2 is due to $\epsilon(s)+1 $ in Eq.~(\ref{eq:35}).

In the presence of the gauge fields $\bar{A}$ and $\bar{B}$, the Chern-Simons form $\omega_5$ is
\begin{eqnarray}
 && \int \omega_5(\bar{A},\bar{B})\label{eq:16} \\
 &=&\int [ ~\{d(\bar{A}+\bar{B})\}^2~(\bar{A}+\bar{B}) + \{d(-\bar{A}+\bar{B})\}^2~(-\bar{A}+\bar{B}) ~] \\
 &=&\int [~ 2(d\bar{A})^2 \bar{B} + 4~ d\bar{A} ~ d\bar{B} ~ \bar{A} + \mathcal{O}(\bar{B}^2)~ ].\label{eq:72}
 \end{eqnarray}
 Note that the $\bar{B}$ dependent part does not vanish although the anomaly-free condition for $\bar{A}$ is satisfied.
 By substituting Eq.~(\ref{eq:72}) into Eq.~(\ref{eq:26}), we obtain
\begin{equation}
 j_\mu^{(CS)}(x,s) = \frac{-1}{4\pi^2} ~\epsilon_{\mu abcd}~\partial_a\bar{A}_b \partial_c\bar{A}_d (x,s) \label{eq:104},
\end{equation}
where $\mu=1,\cdots,4$ because $\bar{B}_5 =0$, and $a,b,c,d=1,\cdots,5$.
Then, the divergence of $J_\mu^{(CS)}$ is calculated as follows:
\begin{eqnarray}
 \partial_\mu J_\mu^{(CS)}(x) &=& \partial_\mu \int^{L}_0 ds~ j_\mu^{(CS)}(x,s) \\
 &=& \frac{-1}{4\pi^2}  \int^{L}_0 ds~ \partial_\mu \left(\epsilon_{\mu abcd}~\partial_a \bar{A}_b \partial_c \bar{A}_d  \right)\\
 &=& \frac{-1}{2\pi^2}  \int^{L}_0 ds~ \epsilon_{\mu 5bcd}~ \left( \partial_\mu \partial_5 \bar{A}_b \right)  \partial_c \bar{A}_d \\
  &=& \frac{1}{4\pi^2}  \int^{L}_0 ds~\partial_5 \left(\epsilon_{5\mu bcd}~\partial_\mu \bar{A}_b \partial_c \bar{A}_d  \right)\\
 &=& \frac{1}{4\pi^2} \left. \epsilon_{\mu\nu\lambda\rho} ~\partial_\mu \bar{A}_\nu \partial_\lambda \bar{A}_\rho \right|_{s=0}^{s=L} \\
  &=& \frac{-1}{16\pi^2} \epsilon_{\mu\nu\lambda\rho}F_{\mu\nu}F_{\lambda\rho}(x,0) ,\label{eq:36}
\end{eqnarray}
with $\mu,\nu,\lambda,\rho = 1,\cdots,4$.
We have used in the last line that $\bar{A}_\mu(x,s=L)$ is pure gauge. 

On the other hand, the anomaly of $J^{(q_L,q_R)}_\mu$ is the same as the conventional axial anomaly of the vector-like fermion\cite{Adler:1969gk,Bell:1969ts}:
\begin{equation}
  \partial_\mu J_\mu^{(q_L,q_R)}(x) = \frac{1}{16\pi^2}  \epsilon_{\mu\nu\lambda\rho}F_{\mu\nu}F_{\lambda\rho}(x,s=0) . \label{eq:1}
\end{equation} 
$\partial_\mu J_\mu^{(\mathrm{fluff})}$ is similar, but vanishes because $\bar{A}(x,s=L)$ is pure gauge\footnote{The fluff fermions are indeed decoupled even for the anomaly.}.
 
Thus the 4-dimensional current $J_\mu^B$ is conserved as mentioned above:
\begin{eqnarray}
 \partial_\mu J_\mu^B(x)&=& \partial_\mu  J_\mu^{(q_L,q_R)}(x) + \partial_\mu J_\mu^{(\mathrm{fluff})} + \partial_\mu J_\mu^{(CS)}(x) \\
 &=& 0 .
\end{eqnarray}
In addition, the current is non-local in the sense of the 4-dimensional field theory because it includes the bulk contribution.
Therefore we can not regard $J_\mu^B$ as the local U(1) axial current in the effective theory.
%$\tilde{j}_\mu$ includes not only an axial current but also a topological current.
%Therefore, from the viewpoint of the 4-dimensional effective theory,

In order to obtain the local and correctly anomalous current, we subtract the bulk contribution from $J_\mu^B$:
\begin{equation}
 J_\mu^{axial}(x)  \equiv J_\mu^B(x) - \int d^4y\int^{L}_0 ds ~ j_\nu^{(CS)}(y,s) ~\frac{\delta \bar{B}_\nu(y,s)}{\delta B_\mu(x)}. \label{eq:11}
\end{equation}
Indeed, Eq.~(\ref{eq:11}) can be written as
\begin{equation}
 J_\mu^{axial} (x)  = \int d^4y \int _0^L ds~ \left( j_\nu^B (y,s) -j_\nu^{(CS)}(y,s) \right) ~\frac{\delta \bar{B}_\nu(y,s)}{\delta B_\mu(x)} \label{eq:15} ,
\end{equation}
%because $J_\mu^B(x)$ can be written 
%  \begin{equation}
%   J_\mu^B(x) = \int d^4y \int _0^L ds~ j_\nu^B (y,s) ~\frac{\delta \bar{B}_\nu(y,s)}{\delta B_\mu(x)} \label{eq:14}.
%  \end{equation}
which is manifestly local and reproduces the correct anomaly:
\begin{equation}
 \partial_\mu J_\mu^{axial} = \frac{1}{16\pi^2}\epsilon_{\mu\nu\lambda\rho}F_{\mu\nu}F_{\lambda\rho}(x) .
\end{equation}
%Interestingly, this procedure is similar to that when one obtains the gauge invariant axial current from the gauge variant conserved current $\tilde{J}_\mu$, which appears in the context of the QCD U(1) problem\cite{tHooft:1986ooh}.
%

Note that the Chern-Simons current $j_\mu^{(CS)}(x,s)$ and $j_\mu^{B}(x,s)$ are gauge invariant (See Eq.~(\ref{eq:104}) and Eq.~(\ref{eq:100})).
%The latter is trivial by the construction.
Therefore $J_\mu^{axial}(x)$ is also gauge invariant.
This is true also when the gauge group of the gauge field $\bar{A}_\mu$ is non-Abelian.
%This is trivial for $j_\mu^{B}(x,s)$.
In such case, indeed, the Chern-Simons form is
\begin{eqnarray}
 && \int \omega_5(\bar{A},\bar{B}) \\
 &=& \sum_{R=r,\bar{r}}\int \mathrm{tr}_R \left[ (d(\bar{A}+\bar{B}))^2 (\bar{A}+\bar{B}) + \frac{3}{2}(\bar{A}+\bar{B})^3 d(\bar{A}+\bar{B}) + \frac{3}{5}(\bar{A}+\bar{B})^5 \right] \label{eq:89},
\end{eqnarray}
where $r$ and $\bar{r}$ are the representations of the two fermions $\psi$ and $\psi'$, respectively.
Thus the Chern-Simons current is written as
\begin{equation}
 j_\mu^{(CS)} (x,s) = \frac{-1}{32\pi^2} \sum_{R=r,\bar{r}}\epsilon_{\mu abcd}~\mathrm{tr}_R ~\bar{F}_{ab}\bar{F}_{cd}~,\label{eq:90}
\end{equation}
which is manifestly gauge invariant.
% the divergence of $J_\mu^{(CS)}$ is calculated as Eq.~(\ref{eq:36}):
%\begin{equation}
% \partial_\mu J_\mu^{(CS)} = \int _0^L ds ~ \frac{-1}{16\pi^2} \sum_{\psi,\psi'}\epsilon_{\mu\nu\lambda\rho} F_{\mu\nu}F_{\lambda\rho}\label{eq:17}.
%\end{equation}
%%%%%%%%
 \section{Summary and conclusions}\label{sec:summary-conclusions}
 \quad
In this paper, we have studied the formulation in Ref.\cite{Grabowska:2015qpk,Grabowska:2016bis} in the continuum.
In Sec.~\ref{sec:regul-form}, we have given the regularization by Eq.~(\ref{fermion_det_reg5}) with Eq.~(\ref{eq:50}) and Eq.~(\ref{eq:45}).
The Pauli-Villars pairs could generate extra massless modes on the walls and Chern-Simons terms in the bulk.
However, the condition, Eq.~(\ref{eq:45}), eliminates these extra contributions.

In Sec.~\ref{sec:calc-doma-wall}, we have calculated the effective action to the quadratic order in the gauge field, and we have found that the effective action consists of three parts.
One is the effective action of the chiral fermions on the domain walls with Pauli-Villars-like regularization.
 The second is the Chern-Simons term in the bulk.
 The third are divergent terms, which are cancelled by the Pauli-Villars pairs.
%We carried out this calculation only to the quadratic order, but we expect that this result is correct to all orders.
%It is confirmed that the regularization argued in Sec.~\ref{sec:regul-form} is successful and that the massless modes localized on the domain walls look like chiral fermions even at the quantum level.

In Sec.~\ref{sec:constr-u1-axial}, we have argued the axial-vector current in 4 dimensions. 
We have introduced two sets of domain-wall fermions belonging to complex conjugate representations so that the effective theory is the vector-like gauge theory.
Then we have considered the axial-vector current that generates the simultaneous phase transformations for the fermions.
This current is exactly conserved, but it contains the contribution from the bulk, which is non-local from the viewpoint of the 4-dimensional theory.
Therefore the local gauge invariant axial-vector current is obtained by subtracting the bulk part.
 \section*{Acknowledgment}\label{sec:acknowledgment}
 \quad
 We would like to thank Hiroshi Suzuki, Okuto Morikawa and Ryuichiro Kitano for useful discussion at KEK Theory Workshop 2016.
%%%%
\begin{appendices}
 
\section{Propagator of domain-wall fermion} \label{appendix}
\quad
The propagator of the domain-wall fermion is a solution of the following equation:
\begin{equation}
[ i \slashed{p}+\gamma^5\partial_s - \epsilon(s)M ] G(p,s;s')=\delta(s-s') , \label{eq:85}
\end{equation}
where $G(p,s;s')$ is the Fourier transform of the propagator in the $2n$ directions,
 \begin{equation}
  G(x,s;x',s') = \int \frac{d^{2n}p}{(2\pi)^{2n}} ~e^{-ip\cdot (x-x')} ~G(p,s;s') .
 \end{equation}
We first consider the region $s'>0$.
 Then we have three cases for $s$:
\begin{equation}
 \left\{
\begin{matrix}
  &\mathrm{(i)}& 0<s'<s \\
  &\mathrm{(ii)}& 0<s<s' \\
 &\mathrm{(iii)}& s<0<s'
 \end{matrix}
  \right.
\end{equation}

We denote the propagators for (i),(ii),(iii) by $G^{(1)}$, $G^{(2)}$, $G^{(3)}$, respectively.
From Eq.~(\ref{eq:85}), we have
%\begin{equation}
% \left[i\slashed{p} + \gamma^5 \partial_s - M \right]G^{(1)}(p,s;s')=0,
%\end{equation}
\begin{equation}
 G^{(1)}(p,s;s') = e^{(i\slashed{p}+M )\gamma^5 s} ~C_1(s'),
\end{equation}
\begin{equation}
G^{(2)}(p,s;s') = e^{(i\slashed{p}+M )\gamma^5 s} ~C_2(s').
\end{equation}
\begin{equation}
 G^{(3)}(p,s;s') = e^{(i\slashed{p}-M )\gamma^5 s} ~C_3(s'),
\end{equation}
where $C_1,C_2,C_3$ are $s$ independent matrices.
Note that the sign of the mass in $G^{(3)}$ is different from the others.
We impose the following boundary conditions:
\begin{equation}
 \begin{cases}
 G^{(1)} (s=s') - G^{(2)}(s=s') = \gamma^5 \\
  G^{(2)}(s=0) = G^{(3)}(s=0) \\
  G^{(1)}(s=L) = G^{(3)}(s=-L) .
 \end{cases}
\end{equation}
The first equation is obtained from Eq.~(\ref{eq:85}) by integrating for $s$ around $s'$.
 The second is to connect $G$ continuously at $s=0$.
 The third is the periodic boundary condition.
Thus matrices $C_1,C_2,C_3$ are all determined.
Then, by using the identity
\[
 e^{(i\slashed{p}+M)\gamma^5 s} = \cosh \left(s\sqrt{p^2+M^2}\right) + \frac{(i\slashed{p}+M)\gamma^5}{\sqrt{p^2+M^2}} \sinh \left(s\sqrt{p^2+M^2}\right) ,
\]
we obtain
\begin{eqnarray}
  G^{(1)} (p,s;s')&=& \frac{-\sqrt{p^2+M^2}}{2\sinh(L\sqrt{p^2+M^2})} ~\frac{i\slashed{p}}{p^2} ~e^{-(i\slashed{p}+M)\gamma^5(s-L)}  ~e^{(i\slashed{p}-M)\gamma^5s'}, \\
 G^{(2)} (p,s;s') &=& \frac{-\sqrt{p^2+M^2}}{2\sinh(L\sqrt{p^2+M^2})} ~\frac{i\slashed{p}}{p^2} ~e^{-(i\slashed{p}+M)\gamma^5s} ~ e^{(i\slashed{p}-M)\gamma^5(s'-L)} , \\
 G^{(3)}(p,s;s') &=& \frac{-\sqrt{p^2+M^2}}{2\sinh(L\sqrt{p^2+M^2})} ~\frac{i\slashed{p}}{p^2} ~e^{-(i\slashed{p}+M)\gamma^5(s-s'+L)} .
\end{eqnarray}

In the following, we consider the limit of $L\rightarrow\infty$.
Then, $G^{(1)}$ becomes
\begin{eqnarray}
 &&  G^{(1)} (p,s;s') \nonumber \\
&=& \frac{i\slashed{p}}{p^2} ~\frac{-\sqrt{p^2+M^2}}{e^{L\sqrt{p^2+M^2}} - e^{-L\sqrt{p^2+M^2}}} \nonumber \\
  && \hspace{2em}\times ~\left[\cosh \{(s-L)\sqrt{p^2+M^2}\} - \frac{(i\slashed{p}+M)\gamma^5}{\sqrt{p^2+M^2}} \sinh\{(s-L)\sqrt{p^2+M^2}\} \right]  ~e^{(i\slashed{p}-M)\gamma^5s'}\nonumber \\
  &\rightarrow& -\frac{i\slashed{p}}{2p^2} ~  e^{-s\sqrt{p^2+M^2}} ~ ~\left[\sqrt{p^2+M^2}+(i\slashed{p}+M)\gamma^5 \right] \nonumber \\
  && \hspace{2em} \times ~\left[\cosh\left( s'\sqrt{p^2+M^2} \right) + \frac{(i\slashed{p}-M)\gamma^5}{\sqrt{p^2+M^2}}\sinh\left( s'\sqrt{p^2+M^2}\right) \right] \label{eq:102}\nonumber  \\
%  &=& \frac{-1}{4}\frac{i\slashed{p}}{p^2} \frac{1}{\sqrt{p^2+M^2}} ~ \left[\sqrt{p^2+M^2}+(i\slashed{p}+M)\gamma^5\right]~ \left[\sqrt{p^2+M^2}+(i\slashed{p}-M)\gamma^5 \right]~ e^{(s'-s)\sqrt{p^2+M^2}} \nonumber \\
% && +\frac{-1}{4}\frac{i\slashed{p}}{p^2} \frac{1}{\sqrt{p^2+M^2}} ~ \left[\sqrt{p^2+M^2}+(i\slashed{p}+M)\gamma^5\right]~ \left[\sqrt{p^2+M^2}+(-i\slashed{p}+M)\gamma^5 \right]~ e^{-(s+s')\sqrt{p^2+M^2}} \\
 &=& -\frac{i\slashed{p}+M-\sqrt{p^2+M^2}~\gamma^5}{2\sqrt{p^2+M^2}}~e^{(s'-s)\sqrt{p^2+M^2}} \nonumber \\
  &&-\frac{i\slashed{p}M(i\slashed{p}+\sqrt{p^2+M^2}~\gamma^5 +M)}{2p^2 \sqrt{p^2+M^2}}e^{-(s+s')\sqrt{p^2+M^2}}.\label{eq:103}
\end{eqnarray}
 $G^{(2)}$ and $G^{(3)}$ can be calculated similarly.
 The result can be summarized as
%
% 
%  S^{(+)} (p,s-s') ~+  &(0<s,s')\\
%	    S^{(-)} (p,s-s') ~+ ~\mathcal{D}^{(-)}(p) ~ e^{(s'+s)\sqrt{p^2+M^2}}  &(s,s'<0)\\
%	    &(s<0<s')\\
%	    \mathcal{D}^{(+-)}(p) ~ e^{(s'-s)\sqrt{p^2+M^2}} &(s'<0<s)
%
%
% 
\begin{equation}
 \begin{cases}
	    G^{(1)}(p,s;s')=  S^{(+)} (p,s-s') +  ~\mathcal{D}^{(+)}(p) ~ e^{-(s'+s)\sqrt{p^2+M^2}} &(0<s'<s)\\
	    G^{(2)}(p,s;s')=  S^{(+)} (p,s-s') +  ~\mathcal{D}^{(+)}(p) ~ e^{-(s'+s)\sqrt{p^2+M^2}} &(0<s<s')\\
	    G^{(3)}(p,s;s')=  \mathcal{D}^{(-+)}(p) ~ e^{(s-s')\sqrt{p^2+M^2}} &(s<0<s'),
	     \end{cases}  \label{eq:6}
\end{equation}
where
\begin{eqnarray}
  S^{(+)} (p,s-s') &=& - \theta(s-s') ~ \frac{i\slashed{p}+M-\sqrt{p^2+M^2}\gamma^5}{2\sqrt{p^2+M^2}} ~ e^{(s'-s)\sqrt{p^2+M^2}} \nonumber \\
 &&  -\theta(s'-s) ~ \frac{i\slashed{p}+M+\sqrt{p^2+M^2}\gamma^5}{2\sqrt{p^2+M^2}} ~ e^{(s-s')\sqrt{p^2+M^2}}  ,\\
\mathcal{D}^{(+)}(p)&=& - \frac{i\slashed{p}M(i\slashed{p} + \sqrt{p^2+M^2} \gamma^5 +M)}{2p^2\sqrt{p^2+M^2}} , \\
\mathcal{D}^{(-+)}(p)&=& - \frac{i\slashed{p} (\sqrt{p^2+M^2}-(i\slashed{p}-M)\gamma^5)}{2p^2} .\label{eq:13}
\end{eqnarray}

The propagator for $s'<0$ is obtained by replacing $M\rightarrow -M$ and $\gamma^5\rightarrow-\gamma^5$ in the above expressions (\ref{eq:6})-(\ref{eq:13}):
 \begin{equation} \label{eq:8}
 \begin{cases}
	    G^{(4)}(p,s;s')=  S^{(-)} ~(p,s-s') + \mathcal{D}^{(-)}(p) ~ e^{(s'+s)\sqrt{p^2+M^2}} &(s<s'<0)\\
	    G^{(5)}(p,s;s')=  S^{(-)} ~(p,s-s') + \mathcal{D}^{(-)}(p) ~ e^{(s'-s)\sqrt{p^2+M^2}} &(s'<s<0')\\
	    G^{(6)}(p,s;s')=  \mathcal{D}^{(+-)}(p) &(s'<0<s) ,
	     \end{cases}
\end{equation}
where
\begin{eqnarray}
 S^{(-)} (p,s-s') &=& - \theta(s-s') ~ \frac{i\slashed{p}-M-\sqrt{p^2+M^2}\gamma^5}{2\sqrt{p^2+M^2}} ~ e^{(s'-s)\sqrt{p^2+M^2}} \nonumber \\
                  &&  -\theta(s'-s) ~ \frac{i\slashed{p}-M+\sqrt{p^2+M^2}\gamma^5}{2\sqrt{p^2+M^2}} ~ e^{(s-s')\sqrt{p^2+M^2}} , \\
\mathcal{D}^{(-)}(p)&=& + \frac{i\slashed{p}M(i\slashed{p} - \sqrt{p^2+M^2} \gamma^5 -M)}{2p^2\sqrt{p^2+M^2}} , \\
\mathcal{D}^{(+-)}(p)&=& - \frac{i\slashed{p} (\sqrt{p^2+M^2}+(i\slashed{p}+M)\gamma^5)}{2p^2}  .
\end{eqnarray}
\section{Vacuum polarization}\label{sec:details-2-point}
\quad
We give the concrete expressions of I, II, III that are defined in Eq.~(\ref{eq:30}), (\ref{eq:20}), (\ref{eq:32}), respectively.
I is given by
\begin{equation}
\mathrm{I} = \iint _{(\mathrm{I})} ~\mathrm{tr}\left[ \bar{A}_\mu (-k,s')\bar{A}_\nu(k,s) \right] ~\int \frac{d^dp}{(2\pi)^d} ~T_{local}^{(-)}(p,p',s,s') ,\label{eq:74}
\end{equation}
where
\begin{eqnarray}
&& T_{local}^{(-)}(p,p',s,s') \nonumber \\
 & =& \mathrm{tr}\left[ \gamma^\mu D^{(-)} \gamma^\nu D'^{(-)} \right] + \mathrm{tr}\left[ \gamma^\mu D^{(-)} \gamma^\nu S'^{(-)} \right] + \mathrm{tr}\left[ \gamma^\mu S^{(-)} \gamma^\nu D'^{(-)} \right] . \label{eq:75}
\end{eqnarray}
Here, $\mathrm{tr}\left[ \gamma^\mu D^{(-)} \gamma^\nu D'^{(-)} \right] $ is calculated as follows:
\begin{eqnarray}
 && \mathrm{tr}\left[ \gamma^\mu D^{(-)}(p,s';s) \gamma^\nu D^{(-)}(p',s;s') \right] \\
 &=& \frac{-M^2 e^{-(s+s')(\sqrt{p^2+M^2}+\sqrt{p'^2+M^2})}}{4p^2p'^2\sqrt{p^2+M^2}\sqrt{p'^2+M^2}} \nonumber \\
 && \hspace{2em} \times\mathrm{tr}\left[ \gamma^\mu ~\slashed{p}(i\slashed{p} - \sqrt{p^2+M^2}\gamma^5 -M) \gamma^\nu~ \slashed{p}'(i\slashed{p}' - \sqrt{p'^2+M^2}\gamma^5 - M)\right] \label{eq:84} \\
 &\equiv& \alpha(p,p') ~e^{(s+s')(\sqrt{p^2+M^2}+\sqrt{p'^2+M^2})}, \label{eq:76}
\end{eqnarray}
where
\begin{eqnarray}
 \alpha(p,p') &=&  \frac{M^2}{4p^2p'^2 \sqrt{p^2+M^2}\sqrt{p'^2+M^2}} \nonumber \\
 && \times \left[ 2^n p^2p'^2~\delta^{\mu\nu} + 2^n (\sqrt{p^2+M^2}\sqrt{p'^2+M^2}+M^2)N^{\mu\nu} \right. \nonumber \\
&& \hspace{5em} -\left.  M(\sqrt{p^2+M^2}+\sqrt{p'^2+M^2}) \mathrm{tr}\left[\gamma^\mu \slashed{p} \gamma^\nu \slashed{p}' \gamma^5 \right] \right], \label{eq:77} 
\end{eqnarray}
and $ N^{\mu\nu} = p\cdot p'\delta^{\mu\nu} - p^\mu p'^\nu - p^\nu p'^\mu $ .
\vspace{1em}

Similarly, $\mathrm{tr}\left[ \gamma^\mu D^{(-)} \gamma^\nu S'^{(-)} \right]$ is given by
\begin{equation}
  \mathrm{tr}\left[ \gamma^\mu D^{(-)} \gamma^\nu S'^{(-)} \right] \equiv  \beta(p,p') ~ e^{(s'+s)\sqrt{p^2+M^2}} ~e^{(s'-s)\sqrt{p'^2+M^2}} ,\label{eq:78}
\end{equation}
where
\begin{eqnarray}
 \beta(p,p') &=& \frac{M}{4p^2\sqrt{p^2+M^2}\sqrt{p'^2+M^2}} \nonumber \\
 && \times\left[ 2^n M (-p^2 \delta^{\mu\nu} + N^{\mu\nu}) -\sqrt{p^2+M^2} ~\mathrm{tr} \left[ \gamma^\mu \slashed{p} \gamma^\nu \slashed{p}' \gamma^5  \right]  \nonumber \right. \\
 && \hspace{5em}+\left. p^2 \sqrt{p'^2+M^2} ~\mathrm{tr} \left[\gamma^\mu \gamma^\nu \gamma^5\right]\right] .\label{eq:79}
\end{eqnarray}
$\mathrm{tr}\left[ \gamma^\mu S^{(-)} \gamma^\nu D'^{(-)} \right]$ is given by
\begin{equation}
 \mathrm{tr}\left[ \gamma^\mu S^{(-)} \gamma^\nu D'^{(-)} \right] \equiv  \gamma(p,p') ~ e^{(s'+s)\sqrt{p'^2+M^2}} ~e^{(s'-s)\sqrt{p^2+M^2}} ,\label{eq:80}
\end{equation}
where
\begin{eqnarray}
\gamma(p,p') &=& \frac{M}{4p'^2\sqrt{p'^2+M^2}\sqrt{p^2+M^2}} \nonumber \\
 && \times\left[ 2^n M(-p'^2 \delta^{\mu\nu} + N^{\mu\nu}) -\sqrt{p'^2+M^2} ~\mathrm{tr} \left[ \gamma^\mu \slashed{p} \gamma^\nu \slashed{p}' \gamma^5  \right]  \nonumber \right. \\
 && \hspace{5em}+\left. p'^2 \sqrt{p^2+M^2} ~\mathrm{tr} \left[\gamma^\mu \gamma^\nu \gamma^5\right]\right] .\label{eq:81}
\end{eqnarray}

Consequently, I is obtained as follows:
 \begin{eqnarray}
 \mathrm{I}&=& ~\mathrm{tr}\left[ A_\mu (-k)A_\nu(k) \right] \nonumber \\
 && \times \int \frac{d^dp}{(2\pi)^d} \left[ -\frac{ 2^n M^2 \delta^{\mu\nu}}{8(p^2+M^2)(p'^2+M^2)} \right. \nonumber \\
&&  \hspace{3.5em}+ \frac{ 2^n M^2(p'^2\sqrt{p'^2+M^2}+p^2\sqrt{p^2+M^2}) N^{\mu\nu} }{8p^2 p'^2(p^2+M^2)(p'^2+M^2)(\sqrt{p^2+M^2}+\sqrt{p'^2+M^2})} \nonumber \\
 && \hspace{3.5em}+\frac{2^n M^2 ( p^2 p'^2 \delta^{\mu\nu} + (\sqrt{p^2+M^2}\sqrt{p'^2+M^2}+M^2 ) N^{\mu\nu})}{8 p^2 p'^2 (\sqrt{p^2+M^2}+\sqrt{p'^2+M^2})^2 \sqrt{p^2+M^2}\sqrt{p'^2+M^2}} \nonumber \\
 &&\hspace{3.5em} - \frac{M(p^2+p'^2+M^2)~\mathrm{tr}\left[\gamma^\mu\slashed{p}\gamma^\nu\slashed{p}'\gamma^5\right]}{8p^2p'^2\sqrt{p^2+M^2}\sqrt{p'^2+M^2}(\sqrt{p^2+M^2}+\sqrt{p'^2+M^2})} \nonumber \\
 &&\hspace{3.5em} \left. +\frac{M(p^2+p'^2+2M^2)~\mathrm{tr}\left[\gamma^\mu\gamma^\nu\gamma^5\right]}{8(p^2+M^2)(p'^2+M^2)(\sqrt{p^2+M^2}+\sqrt{p'^2+M^2})} \right] \label{eq:73}.
 \end{eqnarray}
The last term that includes $\mathrm{tr}\left[\gamma^\mu\gamma^\nu\gamma^5\right]$ will be cancelled with the contribution from the region $\mathrm{I}'$ because the net effect of interchanging $s\leftrightarrow s'$ changes the sign of $\gamma^5$ in $S^{(-)}$.

% and $\gamma^5$ in $\mathrm{tr}[\gamma^\mu\gamma^\nu \gamma^5]$ has come from $S^{(-)}$.
%On the other hand, the 5th line in Eq.~(\ref{eq:73}) will not because that its $\gamma^5$ has come from $\mathcal{D}^{(-)}$.

Similarly, II is given by
\begin{eqnarray}
 \mathrm{II} &=& \iint _{(\mathrm{II})} ~\mathrm{tr}\left[ \bar{A}_\mu (-k,s')\bar{A}_\nu(k,s) \right] \nonumber \\
 &&~\int \frac{d^dp}{(2\pi)^d} ~\left[\mathrm{tr}\left[\gamma^\mu D^{(+-)} \gamma^\nu D'^{(-+)}\right] - \mathrm{tr}\left[ \gamma^\mu S^{(-)}\gamma^\nu S'^{(-)}\right]\right] \\
&=&  \mathrm{tr}[A_\mu(-k) A_\nu(k) ]\nonumber \\
&&\int \frac{d^dp}{(2\pi)^d} \left[ -\frac{2^n M^2\delta^{\mu\nu}}{4\sqrt{p^2+M^2}\sqrt{p'^2+M^2} (\sqrt{p^2+M^2}+\sqrt{p'^2+M^2})^2} \right. \nonumber \\
 && +\frac{2^n M^2 N^{\mu\nu}(p^2+p'^2+M^2+\sqrt{p^2+M^2}\sqrt{p'^2+M^2})} {4p^2p'^2\sqrt{p^2+M^2}\sqrt{p'^2+M^2}(\sqrt{p^2+M^2}+\sqrt{p'^2+M^2})^2} \nonumber \\
 &&- \frac{M~ \mathrm{tr}\left[\gamma^\mu\slashed{p}\gamma^\nu\slashed{p}'\gamma^5 \right]}{4p^2p'^2(\sqrt{p^2+M^2}\sqrt{p'^2+M^2})} \nonumber \\
 &&\left. +\frac{M~\mathrm{tr}[\gamma^\mu\gamma^\nu\gamma^5]}{4\sqrt{p^2+M^2}\sqrt{p'^2+M^2}(\sqrt{p^2+M^2}+\sqrt{p'^2+M^2})} \right].\label{eq:19}
\end{eqnarray}
Again the last term will be cancelled with the contribution from the region $\mathrm{II}'$.
% but the 4th line in Eq.~(\ref{eq:19}), which has come from $\mathcal{D}^{(-+)}$ and $\mathcal{D}^{(+-)}$, will not. 

III is given by
%reflecting I with respect to $s=0$, i,e, replacing the mass of the domain-wall fermion $M\rightarrow-M$ and all $\gamma^5\rightarrow-\gamma^5$ in region I.
\begin{eqnarray}
 \mathrm{III} &=&   \iint _{(\mathrm{III})} ~\mathrm{tr}\left[ \bar{A}_\mu (-k,s')\bar{A}_\nu(k,s) \right] \nonumber \\
 && \hspace{2em} \times\int \frac{d^dp}{(2\pi)^d} \left[ T_{local}^{(+)} (p,p',s,s')+ T_{bulk}(p,p',s,s')\right] , 
\end{eqnarray}
where
\begin{equation}
 T_{bulk} = \mathrm{tr}\left[ \gamma^\mu S^{(+)} \gamma^\nu S'^{(+)} \right]  -\mathrm{tr}\left[ \gamma^\mu S^{(-)} \gamma^\nu S'^{(-)} \right] ,
\end{equation}
\begin{eqnarray}
 T_{local}^{(+)} &=& \mathrm{tr}\left[ \gamma^\mu D^{(+)} \gamma^\nu D'^{(+)} \right] + \mathrm{tr}\left[ \gamma^\mu D^{(+)} \gamma^\nu S'^{(+)} \right] + \mathrm{tr}\left[ \gamma^\mu S^{(+)} \gamma^\nu D'^{(+)} \right] .
% &=& \alpha(p,p')~ e^{-(s+s')(\sqrt{p^2+M^2}+\sqrt{p'^2+M^2})} \nonumber \\
% && +~ \beta(p,p') ~ e^{-(s'+s)\sqrt{p^2+M^2}} ~e^{(s'-s)\sqrt{p'^2+M^2}} \nonumber \\
% && +~ \gamma(p,p') ~e^{-(s'+s)\sqrt{p'^2+M^2}}~e^{(s'-s)\sqrt{p^2+M^2}} 
\end{eqnarray}
Here, $\mathrm{tr}\left[ \gamma^\mu D^{(+)} \gamma^\nu D'^{(+)} \right]$ is calculated similarly to Eq.~(\ref{eq:76}):
\begin{eqnarray}
 && \mathrm{tr}\left[ \gamma^\mu D^{(+)} \gamma^\nu D'^{(+)} \right] \\
 &=& \frac{-M^2 e^{-(s+s')(\sqrt{p^2+M^2}+\sqrt{p'^2+M^2})}}{4p^2p'^2\sqrt{p^2+M^2}\sqrt{p'^2+M^2}} \nonumber \\
 && \hspace{2em} \times\mathrm{tr}\left[ \gamma^\mu ~\slashed{p}(i\slashed{p} + \sqrt{p^2+M^2}\gamma^5 +M) \gamma^\nu~ \slashed{p}'(i\slashed{p}' + \sqrt{p'^2+M^2}\gamma^5 + M)\right] \label{eq:83} \\
 &=& \alpha(p,p') ~e^{-(s+s')(\sqrt{p^2+M^2}+\sqrt{p'^2+M^2})}.\label{eq:86}
\end{eqnarray}
%\begin{eqnarray}
% \alpha(p,p') &=&  \frac{M^2}{4p^2p'^2 \sqrt{p^2+M^2}\sqrt{p'^2+M^2}} \nonumber \\
% && \times \left[ 2^n p^2p'^2~\delta^{\mu\nu} + 2^n (\sqrt{p^2+M^2}\sqrt{p'^2+M^2}+M^2)N^{\mu\nu} \right. \nonumber \\
%&& \hspace{5em} -\left.  M(\sqrt{p^2+M^2}+\sqrt{p'^2+M^2}) \mathrm{tr}\left[\gamma^\mu \slashed{p} \gamma^\nu \slashed{p}' \gamma^5 \right] \right]. \label{eq:82}
%\end{eqnarray}
Note that $\alpha(p,p')$ in Eq.~(\ref{eq:86}) is equal to Eq.~(\ref{eq:77}).
 %although the signs of $M$ and $\gamma^5$ in Eq.~(\ref{eq:83}) are opposite to that in Eq.~(\ref{eq:84}).
We obtain the similar results for $\mathrm{tr}\left[ \gamma^\mu D^{(+)} \gamma^\nu S'^{(+)} \right]$ and $\mathrm{tr}\left[ \gamma^\mu S^{(+)} \gamma^\nu D'^{(+)} \right]$, and $ T_{local}^{(+)} $ is written as Eq.~(\ref{eq:5}).
%\begin{eqnarray}
% T_{local}^{(+)} &=& \alpha(p,p')~ e^{-(s+s')(\sqrt{p^2+M^2}+\sqrt{p'^2+M^2})} \nonumber \\
% && +~ \beta(p,p') ~ e^{-(s'+s)\sqrt{p^2+M^2}} ~e^{(s'-s)\sqrt{p'^2+M^2}} \nonumber \\
% && +~ \gamma(p,p') ~e^{-(s'+s)\sqrt{p'^2+M^2}}~e^{(s'-s)\sqrt{p^2+M^2}} .
%\end{eqnarray}
%Unlike region I, $\mathrm{tr}\left( \gamma^\mu S^{(+)}(p,s';s) \gamma^\nu S^{(+)}(p',s;s') \right)$ and the subtracting field do not completely cancel but generate a Chern-Simons term.
%The other parts are the same as I.
%
%\begin{eqnarray}
% &&\Gamma^{(\mathrm{III})\mu\nu}(k,s,s') \nonumber \\
% &=& \int\frac{d^dp}{(2\pi)^d}\left[ \mathrm{tr}\left( \gamma^\mu G^{(5)}(p,s';s) \gamma^\nu G^{(4)}(p',s;s') \right) - (\mathrm{subtracting~field}) \right] \nonumber \\ 
% &=&  \int\frac{d^dp}{(2\pi)^d}\left[ \mathrm{tr}\left( \gamma^\mu S^{(-;\leftarrow)}(p,s';s) \gamma^\nu S^{(-;\rightarrow)}(p',s;s') \right) + \mathrm{tr}\left( \gamma^\mu S^{(-;\leftarrow)}(p,s';s) \gamma^\nu D^{(-)}(p',s;s') \right) \right. \nonumber \\
% && \hspace{4em}+  \mathrm{tr}\left( \gamma^\mu D^{(-)}(p,s';s) \gamma^\nu S^{(-;\rightarrow)}(p',s;s') \right) + \mathrm{tr}\left( \gamma^\mu D^{(-)}(p,s';s) \gamma^\nu D^{(-)}(p',s;s')\right) \nonumber \\
% && \hspace{4em} \bigl. - (\mathrm{subtracting~field}) \bigr] \nonumber
%\end{eqnarray}
%
\end{appendices}

\end{document}